\documentclass[aps,prd,letterpaper,showkeys,showpacs,twocolumn,nofootinbib,preprintnumbers,longbibliography,unsortedaddress]{revtex4-1}

\usepackage[top=2.8cm, bottom=2.8cm, left=2cm, right=2cm]{geometry}
\usepackage[utf8]{inputenc}  
\usepackage{amsmath,amssymb,lmodern,amsfonts} 
\usepackage{natbib}
\usepackage{hyperref} 
\usepackage{color}
\usepackage{tabulary} 
\usepackage{empheq}
\usepackage[x11names]{xcolor}

\newcommand*\widefcolorbox[1]{\setlength\fboxrule{0.8pt}\fcolorbox{DarkSeaGreen3}{LemonChiffon1!15!}{\hspace{1em}#1\hspace{1em}}}

\def\beq{\begin{equation}}
\def\eeq{\end{equation}}
\def\bi{\begin{itemize}}
\def\ei{\end{itemize}}
\def\ba{\begin{array}}
\def\ea{\end{array}}
\def\bfig{\begin{figure}}
\def\efig{\end{figure}}

\def\tA{\hat \pi} 
\def\B{{\cal B}}

\def\ab{\alpha_1}
\def\aa{\alpha_2}
\def\ac{\alpha_3}
\def\ad{\alpha_4}

\def\An{\pi_*}
\def\dotAn{\dot \pi_*}
\def\A{{\cal A}}
\def\V{{\cal V}}

\newcommand{\eqn}[1]{(\ref{#1})}
\newcommand{\eq}[1]{Eq.~(\ref{#1})}

\begin{document}

\preprint{PI/UAN-2020-678FT}

\title{Generalized SU(2) Proca theory reconstructed and beyond}

\author{Alexander Gallego Cadavid}
\email{alexander.gallego@uv.cl}
\affiliation{Instituto de F\'{\i}sica y Astronom\'{\i}a,  Universidad  de Valpara\'{\i}so, \\ Avenida Gran Breta\~na 1111,  Valpara\'{\i}so  2360102,  Chile}


\author{Yeinzon Rodr\'iguez}
\email{yeinzon.rodriguez@uan.edu.co}
\affiliation{Centro de Investigaciones en Ciencias B\'asicas y Aplicadas, Universidad Antonio Nari\~no, \\ Cra 3 Este \# 47A-15, Bogot\'a D.C. 110231, Colombia}
\affiliation{Escuela  de  F\'isica,  Universidad  Industrial  de  Santander, \\ Ciudad  Universitaria,  Bucaramanga  680002,  Colombia}
\affiliation{Simons Associate at The Abdus Salam International Centre for Theoretical Physics, \\ Strada Costiera 11, I-34151, Trieste, Italy}

\author{L. Gabriel G\'omez}
\email{luis.gomez@correo.uis.edu.co}
\affiliation{Escuela  de  F\'isica,  Universidad  Industrial  de  Santander, \\ Ciudad  Universitaria,  Bucaramanga  680002,  Colombia}


\begin{abstract}
As a modified gravity theory that introduces new gravitational degrees of freedom, the generalized SU(2) Proca theory (GSU2P for short) is the non-Abelian version of the well-known generalized Proca theory where the action is invariant under global transformations of the SU(2) group.  This theory was formulated for the first time in {\it Phys. Rev. D} {\bf 94} (2016) 084041, having implemented the required primary constraint-enforcing relation to make the Lagrangian degenerate and remove one degree of freedom from the vector field in accordance with the irreducible representations of the Poincar\'e group.  It was later shown in {\it Phys. Rev. D} {\bf 101} (2020) 045008, ibid 045009, that a secondary constraint-enforcing relation, which trivializes for the generalized Proca theory but not for the SU(2) version, was needed to close the constraint algebra.  It is the purpose of this paper to implement this secondary constraint-enforcing relation in GSU2P and to make the construction of the theory more transparent.  Since several terms in the Lagrangian were dismissed in {\it Phys. Rev. D} {\bf 94} (2016) 084041 via their equivalence to other terms through total derivatives, not all of the latter satisfying the secondary constraint-enforcing relation, the work was not so simple as directly applying this relation to the resultant Lagrangian pieces of the old theory.  Thus, we were motivated to reconstruct the theory from scratch.  In the process, we found the beyond GSU2P.
\end{abstract}



\maketitle

\section{Introduction} \label{Intro}

Whether a classical description of the gravitational interaction is fundamental or effective remains a mystery.  What 
is certain is that, no matter whether the fundamental theory of gravity is classical or quantum, and despite its enormous experimental success \cite{will,Psaltis:2020lvx,Akiyama:2019cqa,TheLIGOScientific:2017qsa,GBM:2017lvd,Goldstein:2017mmi,Abuter:2018drb,Collett:2018gpf,Ezquiaga:2018btd,He:2018oai,Do:2019txf,Abbott:2018lct,Ishak:2018his}, Einstein's theory of gravity is an effective theory \cite{Kostelecky:2020hbb,Burgess:2003jk,Donoghue:1994dn}.
The inevitable presence of singularities in General Relativity (GR) \cite{Penrose:1964wq,Hawking:1969sw}, even assuming the validity of the cosmic censorship conjecture \cite{Penrose:1969pc,Penrose:1900mp,Penrose:1994}, points to a breakdown of the theory.  
Should the breakdown take place in the infrared, the new theory that encompasses GR might give us some insight about the true nature of the current accelerated expansion of the universe.  The breakdown might take place in the ultraviolet, helping solve the renormalizability problems of GR and illuminating the way to a quantum description of gravity.  Of course, the breakdown might take place in both the infrared and the ultraviolet.  Another option is at an intermediate scale, in the strong gravity regime, which is particularly interesting because the very young multi-messenger astronomy is giving us, and will continue doing it, valuable information about the behaviour of gravity at the scales associated to compact objects such as black holes and neutron stars\footnote{At these scales, however, there might be some contributions from the ultraviolet-complete theory. This means that the regime of validity of the modified gravity theory must also be ensured before applying constraints that belong to other scales or frequencies \cite{deRham:2018red}.}.  We might, therefore, be on the verge of a scientific crisis and a new revolution in Physics, in the sense of Kuhn \cite{kuhn}.

Over the years, several approaches have been proposed to classically extend Einstein's theory of gravity (see Ref. \cite{Heisenberg:2018vsk} for a review).  Perhaps the simplest one, at least in its conception, is giving mass to the gravitational carrier \cite{deRham:2014zqa};  nevertheless, starting from the Fierz-Pauli action \cite{Fierz:1939ix} and arriving to the de Rham-Gabadadze-Tolley (dRGT) ghost-free massive gravity \cite{deRham:2010kj}, the introduction of a massive graviton has shown to be a difficult challenge.  Another possibility is adding space dimensions while preserving the second-order differential structure of the field equations and keeping untouched the gravitational degrees of freedom;  this is the proposal derived from the Lovelock programme \cite{Lovelock:1971yv,Lovelock:1972vz}, as the only curvature invariant that satisfies these requirements in four space-time dimensions is the Einstein-Hilbert term.  A third alternative is invoking new gravitational degrees of freedom, the simplest of them being a scalar field;  the first proposal in this regard was the well-known Brans-Dicke theory \cite{Brans:1961sx}, but this has turned out to be just a particular case of a whole family of Lagrangians that comprise the, nowadays very famous, Horndeski theory \cite{Horndeski:1974wa,Nicolis:2008in,Deffayet:2009mn,Deffayet:2009wt,Deffayet:2011gz,Kobayashi:2011nu,Deffayet:2013lga,Kobayashi:2019hrl}.  The purpose of preserving the second-order differential structure of the field equations is to remove the Ostrogradski ghost \cite{Ostrogradsky:1850fid,Woodard:2006nt,Woodard:2015zca,Ganz:2020skf} that makes the ground state unstable in the presence of interactions.  Notwithstanding, this is not the only way to remove the Ostrogradski ghost, although it is the most transparent;  the degeneracy of the kinetic matrix associated to the degrees of freedom of the theory can be invoked so that primary constraints among the phase space variables are generated \cite{Langlois:2015cwa} -- in this way, the unwanted degrees of freedom can be removed \cite{Langlois:2015skt} even when the differential structure of the field equations is higher order.  This idea was put in action with the introduction of the beyond Horndeski theory \cite{Gleyzes:2014dya,Zumalacarregui:2013pma} and later generalized to what is now known as the degenerate higher-order scalar-tensor theory (DHOST) \cite{Achour:2016rkg,BenAchour:2016fzp,Crisostomi:2016czh}, where a plethora of Lagrangians rose up to the surface.  The application of this idea to the Lovelock programme has, nonetheless, not been fruitful \cite{Crisostomi:2017ugk}, which is, paradoxically, very suggestive.  A fourth alternative is considering other geometric formulations of gravity, i.e., considering not only the curvature but also the torsion and the non metricity as the protagonist geometric objects in the description of the gravitational interaction \cite{Heisenberg:2018vsk,Blagojevic:2002du,Blagojevic:2013xpa,BeltranJimenez:2019tjy}.  This has a long history starting from the Einstein-Cartan theory \cite{Cartan:1923zea,Cartan:1924yea}, which involves curvature and torsion but leaving aside the non metricity, to the coincident gravity proposal \cite{BeltranJimenez:2017tkd}, where the non metricity is the sole protagonist.  Of course, there are more possibilities, some of them with remanent harmless ghosts, they being, therefore, effective theories.

The introduction of new gravitational degrees of freedom has not been kept only in the realm of a scalar field.  Multiple scalar fields have been considered in what are called the multi-Galileon theories \cite{Padilla:2012dx,Sivanesan:2013tba,Padilla:2010ir,Allys:2016hfl}.  More tensor fields can be considered as well, as in the bimetric theory \cite{Hassan:2011zd} which introduces an extra spin-two metric.  The introduction of vector fields \cite{Tasinato:2014eka,Heisenberg:2014rta,Allys:2015sht,Jimenez:2016isa,Allys:2016jaq} and p-forms \cite{Deffayet:2010zh,Almeida:2018fwe,Almeida:2020lsn} has also been investigated.  Even the mixture of a scalar and a vector field, together with gravity, has been explored \cite{Heisenberg:2018acv}.  Each one of these proposals has its own motivations, which we will not describe here except for those related to the introduction of vector fields.  

The most frequent question when we speak about vector fields in gravity and/or cosmology is:  why to introduce them?  We think the right question is:  why not?:  at the end of the day, and being pragmatical, we have observed many more vector fields in nature than fundamental scalar fields.  We have to be careful with the problems they can generate:  ghosts, anisotropies in cosmology, etc., but this does not preclude their study.  In fact, the role of vector fields in gravitation, astrophysics, and cosmology has attracted a lot of interest in recent years (see Refs. \cite{Heisenberg:2018vsk,Dimopoulos:2011ws,Maleknejad:2012fw,Soda:2012zm} for some reviews), culminating in the construction and study of what is called the generalized Proca theory \cite{Tasinato:2014eka,Heisenberg:2014rta,Allys:2015sht,Jimenez:2016isa,Allys:2016jaq}.  This is the Proca theory \cite{Proca:1900nv,Proca:1939}, in curved spacetime, devoid of internal gauge symmetries and can be seen as the vector-field version of the Horndeski theory\footnote{For the U(1) gauge-invariant version of the generalized Proca theory in flat spacetime see Ref. \cite{Deffayet:2013tca} and in curved spacetime see Ref. \cite{Horndeski:1976gi}.}.  By construction, it is plainly degenerate in order to avoid the propagation of a fourth degree of freedom which clearly disagrees with the structure of the irreducible representations of the Poincar\'e group.  Its decoupling limit, in contrast, reduces to the Horndeski theory.

The generalized Proca theory has been well studied in astrophysics and cosmology \cite{Tasinato:2014eka,DeFelice:2016cri,DeFelice:2016yws,DeFelice:2016uil,Heisenberg:2016wtr,deFelice:2017paw,Heisenberg:2020xak,Heisenberg:2017hwb,Kase:2017egk,Kase:2018voo,Kase:2020yhw}.  In the latter, however, special attention has been paid because of the anisotropies that a vector field produces, inherent to its nature, both in the expansion of the universe and in the cosmological perturbations.  Such anisotropies can easily go beyond the observational constraints, so it is necessary to take some measures such as the rapid oscillations of the vector field \cite{Cembranos:2012kk}, the dilution of the vector field by a companion scalar field \cite{Watanabe:2009ct}, the suppression of the spatial components of the vector field against its temporal component (what is called the temporal gauge setup) \cite{DeFelice:2016yws}, or the implementation of a cosmic triad of vector fields\footnote{The cosmic triad is a set of three vector fields mutually orthogonal and of the same norm.} that restores the isotropy \cite{ArmendarizPicon:2004pm,Golovnev:2008cf,Emami:2016ldl,Alvarez:2019ues,Gomez:2020sfz}.  The latter proposal has been investigated in different contexts and finds a natural home in the presence of an internal SU(2) symmetry \cite{Maleknejad:2011jw,Adshead:2012kp,Nieto:2016gnp,Adshead:2017hnc,Guarnizo:2020pkj}.  Indeed, the temporal gauge setup and the cosmic triad are two of the four possible setups that are compatible with a spatial spherical symmetry and that are realized under an internal SU(2) symmetry \cite{Witten:1976ck,Sivers:1986kq,Forgacs:1979zs}. This was the main motivation behind the formulation of what was baptized as the generalized SU(2) Proca theory (GSU2P for short) \cite{Allys:2016kbq} (see also Ref. \cite{Jimenez:2016upj}). 
The possible setups mentioned above
spontaneously break the internal (global) SU(2) symmetry along with the Lorentz rotational symmetry and Lorentz boosts, leaving, however, a diagonal spatial rotation subgroup unbroken.  The isotropic expansion of the universe can then be naturally modeled with 
any of the four setups or linear combinations of them
without resorting to fast oscillations or other (scalar) fields.
The price to pay, however, which is the spontaneous breaking of the Lorentz invariance, is, anyway, extraordinarily reasonable, since this seems to be nature's strategy to produce all the patterns we see in condensed matter systems (fluids, superfluids, solids, and supersolids; see Ref. \cite{Nicolis:2015sra}).  Indeed, according to the pattern classification in Ref. \cite{Nicolis:2015sra}, what would be the condensed matter analogs of the temporal gauge setup and the cosmic triad in the GSU2P are the, yet unobserved, type-I and type-II framids, repectively.  The application of the GSU2P to dark energy and inflation has been explored in Refs. \cite{Rodriguez:2017ckc,Rodriguez:2017wkg} and its stability properties in Ref. \cite{Gomez:2019tbj}.

The GSU2P was built in Ref. \cite{Allys:2016kbq} (see also Ref. \cite{Jimenez:2016upj}) having in mind the primary constraints required to remove the fourth degree of freedom\footnote{Concretely, the temporal component of the vector field.}.  To that end, a primary constraint-enforcing relation related to the primary Hessian of the system was employed.  This was done in flat spacetime following the standard procedure of later covariantizing not before having removed redundant terms in the obtained action via total derivatives.  Later on, two caveats were recognized.  First, the constraint algebra was not closed only with the primary constraints, at least for theories involving more than one vector field \cite{ErrastiDiez:2019ttn,ErrastiDiez:2019trb}\footnote{The constraint algebra of the generalized Proca theory, the latter being a theory that involves just one vector field, turned out to be trivially closed.};  a secondary constraint was identified that closed the constraint algebra and that, therefore, pointed out to the existence of ghosts in the GSU2P.  Second, the redundant terms in flat spacetime turned out to be not necessarily redundant in curved spacetime, which would lead, for sure, to new terms not uncovered in Ref. \cite{Allys:2016kbq};  indeed, such a remark led two of us to rediscover the beyond Proca terms in Ref. \cite{GallegoCadavid:2019zke}, they being the vector analogous of the beyond Horndeski terms, already obtained in Ref. \cite{Heisenberg:2016eld}.  Reformulating the GSU2P in order to implement the secondary constraint-enforcing relation seemed at first sight very easy, because it was a matter of applying this relation to the ``old'' GSU2P and seeing what the result would be.  However, this turned out to be impractical, since many terms had disappeared when employing the total derivatives.  Moreover, the total derivatives employed satisfied the primary constraint-enforcing relation but not necessarily the secondary one, so repairing the old theory quickly became quite a big deal and, therefore, unworthy.  The purpose of this paper is to build from scratch the GSU2P, paying attention to the two caveats already mentioned and following a style of construction based on the decomposition of a first-order derivative $\partial_\mu A_\nu^a$ of the vector field $A_\mu^a$ into its symmetric, 
\begin{equation}
S_{\mu \nu}^a \equiv \partial_\mu A_\nu^a + \partial_\nu A_\mu^a \,,
\end{equation}
and antisymmetric part, 
\begin{equation}
A_{\mu \nu}^a \equiv \partial_\mu A_\nu^a - \partial_\nu A_\mu^a \,.
\end{equation}
Employing this decomposition will simplify things and allow us to deal with a lower number of Lagrangian building blocks as compared with Ref. \cite{Allys:2016kbq}.  In the process, we will find the beyond GSU2P.

The layout of the paper is the following.  In the Section \ref{requirements}, we will enumerate the requirements for the construction of the GSU2P.  In Section \ref{L2}, we will show how an arbitrary function of $A_{\mu \nu}^a$ and $A_\mu^a$ satisfies both the primary and secondary constraint-enforcing relations, leaving only the work of finding the right terms in the action involving at least one $S_{\mu \nu}^a$.  In section \ref{32}, we build the Lagrangian involving one derivative and two vector fields.  Similar procedures are followed in Sections \ref{34}, \ref{40}, \ref{42}, and \ref{50}, where we obtain the Lagrangians involving one derivative and four vector fields, two derivatives only, two derivatives and two vector fields, and three derivatives only, respectively.  In all these cases, the number of space-time indices in the Lagrangian building blocks before contractions with the primitive invariants of the Poincar\'e group is less than or equal to six.  We prefer to keep the construction of the theory up to this level since, as shown in Ref. \cite{Allys:2016kbq}, the number of Lagrangian building blocks we have to consider scales very fast when more space-time indices are considered.  Finally, in Sections \ref{compold} and \ref{compgpt}, we compare the ``new'' or ``reconstructed'' GSU2P with the old GSU2P and with the generalized Proca theory, respectively.  Section \ref{conc} is devoted to the conclusions.  Throughout the text, Greek indices are space-time indices and run from 0 to 3, while Latin indices are internal SU(2) group indices and run from 1 to 3.  The sign convention is the (+++) according to Misner, Thorne, and Wheeler \cite{Misner:1974qy}.

\section{Requirements for the construction of the theory} \label{requirements}

The GSU2P must be built having in mind the following criteria:
\begin{enumerate}
\item The action must be, locally, Lorentz invariant (although the symmetry may be non-linearly realized).
\item The vector field must transform as the adjoint representation of the global transformations belonging to the SU(2) group \cite{Fuchs:1997jv,Ramond:2010zz,Feger:2012bs}.  Accordingly, the action must be invariant under these transformations.
\item The primary constraint-enforcing relation $\mathcal{H}^{0 \nu}_{ab} = 0$, where
\begin{equation}
\mathcal{H}^{\mu \nu}_{ab} \equiv \frac{\partial^2 \mathcal{L}}{\partial \dot{A}^a_\mu \partial \dot{A}^b_\nu} \,,
\end{equation}
is the ``primary'' Hessian and a dot means a time derivative, must be satisfied in flat spacetime in order to make the Lagrangian degenerate.  This is a necessary condition to remove the unwanted degree of freedom \cite{Tasinato:2014eka,Heisenberg:2014rta}.
\item The secondary constraint-enforcing relation $\tilde{\mathcal{H}}^{00}_{ab} = 0$, where
\begin{equation}
\tilde{\mathcal{H}}^{\mu \nu}_{ab} \equiv \frac{\partial^2 \mathcal{L}}{\partial \dot{A}^{[a}_\mu \partial {A}^{b]}_\nu} \,,
\end{equation}
is the ``secondary Hessian'' and the brackets mean unnormalized antisymmetrization,
must be satisfied in flat spacetime so that the primary constraint holds at all times\footnote{This condition bears a great resemblance to that obtained in Refs. \cite{Motohashi:2016ftl,Klein:2016aiq} for mechanical systems with multiple degrees of freedom.}.  This condition together with the preceding one are necessary and sufficient to remove the unwanted degree of freedom in flat spacetime \cite{ErrastiDiez:2019ttn,ErrastiDiez:2019trb}.
\item The decoupling limit of the theory must be free of the Ostrogradski ghost as must happen since the full theory is free of it.  This implies that the scalar limit of GSU2P must belong to the non-Abelian extension of the multi-Galileon theory \cite{Padilla:2012dx,Sivanesan:2013tba,Padilla:2010ir,Allys:2016kbq,Allys:2016hfl} or any of its beyond or DHOST versions.
\end{enumerate}

\section{$\mathcal{L}_2$} \label{L2}

All the Lagrangian pieces $\mathcal{L}_i^A$ built exclusively from contractions of $A_{\mu \nu}^a$ and $A_\mu^a$ with the primitive invariants of the Lorentz group\footnote{They may, of course, either preserve or violate parity.} \cite{Fuchs:1997jv,Ramond:2010zz,Feger:2012bs}, collected in a generic Lagrangian piece called $\mathcal{L}_2 (A_{\mu \nu}^a,A_\mu^a)$, satisfy automatically both the primary and secondary constraint-enforcing relations thanks to the antisymmetry of $A_{\mu \nu}^a$.  To see it, let us calculate the primary and secondary Hessians.  First of all,
\begin{eqnarray}
\frac{\partial \mathcal{L}_i^A}{\partial \dot{A}_\mu^a} &=& \frac{\partial \mathcal{L}_i^A}{\partial A_{\rho \sigma}^c} \frac{\partial A_{\rho \sigma}^c}{\partial \dot{A}_\mu^a} = \frac{\partial \mathcal{L}_i^A}{\partial A_{\rho \sigma}^c} \delta^0_{[\rho} \delta^\mu_{\sigma]} \delta^c_a \nonumber \\
&=& \frac{\partial \mathcal{L}_i^A}{\partial A_{\rho \mu}^a} \Bigr|_{\rho = 0} - \frac{\partial \mathcal{L}_i^A}{\partial A_{\mu \sigma}^a} \Bigr|_{\sigma = 0} = 2 \frac{\partial \mathcal{L}_i^A}{\partial A_{\rho \mu}^a} \Bigr|_{\rho = 0} \,. \label{firstdev}
\end{eqnarray}
Any possible ambiguity in the second line of the previous equation is clarified having in mind that $\mathcal{L}_i^A$ is always written as $A_{\mu \nu}^a$ contracted with an antisymmetric tensor\footnote{Except for the case where no $A_{\mu \nu}^a$ tensors are involved.  However, in such a case, $\frac{\partial \mathcal{L}_i^A}{\partial \dot{A}_\mu^a} = 0$ automatically.}.  Thus,
\begin{eqnarray}
\frac{\partial^2 \mathcal{L}_i^A}{\partial \dot{A}_\nu^b \partial \dot{A}_\mu^a} &=& 2 \frac{\partial^2 \mathcal{L}_i^A}{\partial A_{\rho \mu}^a \partial A_{\alpha \beta}^c} \Bigr|_{\rho = 0} \frac{\partial A_{\alpha \beta}^c}{\partial \dot{A}_\nu^b}  \nonumber \\ 
&=& 2 \frac{\partial^2 \mathcal{L}_i^A}{\partial A_{\rho \mu}^a \partial A_{\alpha \beta}^c} \Bigr|_{\rho = 0} \delta^0_{[\alpha} \delta^\nu_{\beta]} \delta^c_b \nonumber \\
&=& 2 \frac{\partial^2 \mathcal{L}_i^A}{\partial A_{\rho \mu}^a \partial A_{\alpha \nu}^b} \Bigr|_{\rho = 0, \alpha=0} - 2 \frac{\partial^2 \mathcal{L}_i^A}{\partial A_{\rho \mu}^a \partial A_{\nu \beta}^b} \Bigr|_{\rho = 0, \beta=0} \nonumber \\
&=& 4 \frac{\partial^2 \mathcal{L}_i^A}{\partial A_{\rho \mu}^a \partial A_{\alpha \nu}^b} \Bigr|_{\rho = 0, \alpha=0} \,.
\end{eqnarray}
The primary constraint-enforcing relation is, therefore, satisfied:
\begin{equation}
\mathcal{H}^{0 \nu}_{ab} = 4 \frac{\partial^2 \mathcal{L}_i^A}{\partial A_{\rho \mu}^a \partial A_{\alpha \nu}^b} \Bigr|_{\rho = 0, \mu=0, \alpha=0} = 0 \,,
\end{equation}
because of the antisymmetry of $A_{\mu \nu}^a$.

Regarding the secondary constraint-enforcing relation, we obtain from Eq. (\ref{firstdev})
\begin{equation}
\frac{\partial^2 \mathcal{L}_i^A}{\partial A_\nu^b \partial \dot{A}_\mu^a} = 2 \frac{\partial^2 \mathcal{L}_i^A}{\partial A_{\rho \mu}^a \partial A_\nu^b} \Bigr|_{\rho = 0} \,,
\end{equation}
which leads to the secondary Hessian
\begin{equation}
\tilde{\mathcal{H}}^{00}_{ab} =  2 \frac{\partial^2 \mathcal{L}_i^A}{\partial A_{\rho \mu}^{[a} \partial A_{\nu}^{b]}} \Bigr|_{\rho = 0, \mu=0, \nu=0} = 0 \,,
\end{equation}
in view, again, of the antisymmetry of $A_{\mu \nu}^a$.

Hence, we can conclude that the $\mathcal{L}_2 (A_{\mu \nu}^a,A_\mu^a)$ Lagrangian piece satisfies automatically the first and secondary constraint-enforcing relations necessary to propagate only three degrees of freedom.  This is the reason why such a Lagrangian piece is so particular, differing in its structure and arbitrariness from the other Lagrangian pieces we are going to describe in the following.  On the other hand, the generalization of $\mathcal{L}_2$ to curved spacetime is straightforward.

\section{One derivative and two vector fields} \label{32}


Lagrangian building blocks constructed from one derivative and two vector fields, linearly independent from $\mathcal{L}_2$, are terms of the form $S_{\mu \nu}A_{\rho}A_{\sigma}$ which, as can be seen, involve four space-time indices.  Group theory tells us that four building blocks can be constructed upon contractions of $S_{\mu \nu}A_{\rho}A_{\sigma}$ with the following tensors \cite{Fuchs:1997jv,Ramond:2010zz,Feger:2012bs}:
\begin{eqnarray}
&&g^{\mu \nu} g^{\rho \sigma} \,, \nonumber \\
&&g^{\mu \rho} g^{\nu \sigma} \,, \nonumber \\
&&g^{\mu \sigma} g^{\nu \rho} \,, \nonumber \\
&&\epsilon^{\mu \nu \rho \sigma} \,, \label{four}
\end{eqnarray}
where $g^{\mu \nu}$ is the contravariant Minkowski metric and $\epsilon^{\mu \nu \rho \sigma}$ is the Levi-Civita tensor.
Thus, the only building blocks either different than zero or with the potential of becoming different than zero after adding the internal group indices are the following:
\begin{eqnarray}
&&S^\mu_\mu (A \cdot A) \,, \nonumber \\
&&S_{\mu \nu} A^\mu A^\nu\,. 
\end{eqnarray}
The addition of the internal group indices leads to terms of the form $S^a A^b A^c$ that involve three internal group indices and which, from group theory \cite{Fuchs:1997jv,Ramond:2010zz,Feger:2012bs}, can be contracted only with the totally antisymmetric tensor $\epsilon_{a b c}$:\footnote{This tensor represents the structure constants of the SU(2) group.  See, in particular, the Misner, Thorne, and Wheeler treatise on gravitation \cite{Misner:1974qy} for a description of the SU(2) group as a manifold endowed with a metric $g_{ab}$ and an orientability form described by $\epsilon_{a b c}$.}
\begin{eqnarray}
&&S^{\mu a}_\mu (A^b \cdot A^c) \epsilon_{a b c} \,, \nonumber \\
&&S^a_{\mu \nu} A^{\mu b} A^{\nu c} \epsilon_{a b c} \,. 
\end{eqnarray}
Such terms vanish because of the antisymmetry of $\epsilon_{a b c}$, so we conclude that there do not exist terms in GSU2P, linearly independent of $\mathcal{L}_2$, that involve one derivative and two vector fields.

\section{One derivative and four vector fields} \label{34}

\subsection{The Lagrangian building blocks}
Lagrangian building blocks built from one derivative and four vector fields, linearly independent of $\mathcal{L}_2$, are terms of the form $S_{\mu \nu} A_\rho A_\sigma A_\alpha A_\beta$ that involve six space-time indices.  Group theory \cite{Fuchs:1997jv,Ramond:2010zz,Feger:2012bs} tells us that, in this case, the building blocks are constructed upon contractions of $S_{\mu \nu} A_\rho A_\sigma A_\alpha A_\beta$ with the following fifteen permutations of the product of three space-time metrics:
\begin{eqnarray}
&&g^{\mu \nu} g^{\rho \sigma} g^{\alpha \beta} \,, \nonumber \\
&&g^{\mu \nu} g^{\rho \alpha} g^{\sigma \beta} \,, \nonumber \\
&&g^{\mu \nu} g^{\rho \beta} g^{\sigma \alpha} \,, \nonumber \\
&&g^{\mu \rho} g^{\nu \sigma} g^{\alpha \beta} \,, \nonumber \\
&&g^{\mu \rho} g^{\nu \alpha} g^{\sigma \beta} \,, \nonumber \\
&&g^{\mu \rho} g^{\nu \beta} g^{\sigma \alpha} \,, \nonumber \\
&&g^{\mu \sigma} g^{\nu \rho} g^{\alpha \beta} \,, \nonumber \\
&&g^{\mu \sigma} g^{\nu \alpha} g^{\rho \beta} \,, \nonumber \\
&&g^{\mu \sigma} g^{\nu \beta} g^{\rho \alpha} \,, \nonumber \\
&&g^{\mu \alpha} g^{\nu \rho} g^{\sigma \beta} \,, \nonumber \\
&&g^{\mu \alpha} g^{\nu \sigma} g^{\rho \beta} \,, \nonumber \\
&&g^{\mu \alpha} g^{\nu \beta} g^{\rho \sigma} \,, \nonumber \\
&&g^{\mu \beta} g^{\nu \rho} g^{\sigma \alpha} \,, \nonumber \\
&&g^{\mu \beta} g^{\nu \sigma} g^{\rho \alpha} \,, \nonumber \\
&&g^{\mu \beta} g^{\nu \alpha} g^{\rho \sigma} \,, \label{sixg}
\end{eqnarray}
as well as with the following ten products of a space-time metric and a Levi-Civita tensor:
\begin{eqnarray}
&&g^{\nu \rho} \epsilon^{\mu \sigma \alpha \beta} \,, \nonumber \\
&&g^{\nu \sigma} \epsilon^{\mu \rho \alpha \beta} \,, \nonumber \\
&&g^{\nu \alpha} \epsilon^{\mu \rho \sigma \beta} \,, \nonumber \\
&&g^{\nu \beta} \epsilon^{\mu \rho \sigma \alpha} \,, \nonumber \\
&&g^{\rho \sigma} \epsilon^{\mu \nu \alpha \beta} \,, \nonumber \\
&&g^{\rho \alpha} \epsilon^{\mu \nu \sigma \beta} \,, \nonumber \\
&&g^{\rho \beta} \epsilon^{\mu \nu \sigma \alpha} \,, \nonumber \\
&&g^{\sigma \alpha} \epsilon^{\mu \nu \rho \beta} \,, \nonumber \\
&&g^{\sigma \beta} \epsilon^{\mu \nu \rho \alpha} \,, \nonumber \\
&&g^{\alpha \beta} \epsilon^{\mu \nu \rho \sigma} \,. \label{sixge}
\end{eqnarray}
Other five contractions of the form $g \epsilon$ are possible, but they are not linearly independent because of the property:
\begin{eqnarray}
g^{\mu \nu} \epsilon^{\rho \sigma \alpha \beta} &=& g^{\nu \rho} \epsilon^{\mu \sigma \alpha \beta} - g^{\nu \sigma} \epsilon^{\mu \rho \alpha \beta} \nonumber \\
&& + g^{\nu \alpha} \epsilon^{\mu \rho \sigma \beta} - g^{\nu \beta} \epsilon^{\mu \rho \sigma \alpha} \,.
\end{eqnarray}
Thus, only three building blocks either are non vanishing or have the potential of becoming different than zero once the internal group indices are added:\footnote{From now on, the starred Lagrangian building blocks and total derivatives will be those that vanish according to the Poincar\'e group but that otherwise survive when considering also the SU(2) group.}
\begin{eqnarray}
&&S_\mu^\mu (A \cdot A) (A \cdot A) \,, \nonumber \\
&&S_{\mu \nu} A^\mu A^\nu (A \cdot A) \,, \nonumber \\
&&S_{\mu \nu} A^\nu A_\sigma A_\alpha A_\beta \epsilon^{\mu \sigma \alpha \beta} \,.   \;\;\;  (\ast)
\end{eqnarray}
When adding the internal group indices, these terms acquire the form $S^a A^b A^c A^d A^e$ which can be contracted, according to group theory \cite{Fuchs:1997jv,Ramond:2010zz,Feger:2012bs}, only with the following six products of an internal group metric and the respective structure constants:
\begin{eqnarray}
&&g_{ab} \epsilon_{cde} \,, \nonumber \\
&&g_{ac} \epsilon_{bde} \,, \nonumber \\
&&g_{ad} \epsilon_{bce} \,, \nonumber \\
&&g_{bc} \epsilon_{ade} \,, \nonumber \\
&&g_{bd} \epsilon_{ace} \,, \nonumber \\
&&g_{cd} \epsilon_{abe} \,. \label{gfive}
\end{eqnarray}
Other four contractions of the form $g \epsilon$ are possible, but they are not linearly independent because of the property:
\begin{equation}
g_{a e} \epsilon_{bcd} = g_{ab} \epsilon_{cde} - g_{ac} \epsilon_{bde} + g_{ad} \epsilon_{bce} \,.
\end{equation}
Therefore, there exist only four linearly independent building blocks in GSU2P that involve one derivative and four vector fields:
\begin{eqnarray}
\mathcal{L}_3^1 &=& S_{\mu \nu}^a A^{\mu b} A^{\nu c} (A_b \cdot A^e) \epsilon_{ace} \,, \nonumber \\
\mathcal{L}_3^2 &=& S_{\mu \nu}^a A^\nu_a A_\sigma^c A_\alpha^d A_\beta^e \epsilon^{\mu \sigma \alpha \beta} \epsilon_{cde} \,, \nonumber \\
\mathcal{L}_3^3 &=& S_{\mu \nu}^a A_{\sigma a} A^{\nu b}  A_\alpha^d A_\beta^e \epsilon^{\mu \sigma \alpha \beta} \epsilon_{bde} \,, \nonumber \\
\mathcal{L}_3^4 &=& S_{\mu \nu}^a A^{\nu b} A_{\sigma b} A_\alpha^d A_\beta^e \epsilon^{\mu \sigma \alpha \beta} \epsilon_{ade} \,. \label{L14}
\end{eqnarray}

\subsection{The Hessian constraints} \label{Hc14}

The Lagrangian is, hence, written as a linear combination of the Lagrangian building blocks of Eq. (\ref{L14}):
\begin{equation}
\mathcal{L} = \sum_{i=1}^4 x_i \mathcal{L}_3^i \,,
\end{equation}
where the $x_i$ are arbitrary constants.
Because only one derivative has been considered, the primary constraint-enforcing relation is satisfied automatically.  Regarding the secondary constraint-enforcing relation, the secondary Hessian gives the following result:
\begin{eqnarray}
\tilde{\mathcal{H}}^{00}_{ab} &=& 2[-A^{0 c}(A_{[b} \cdot A^e) \epsilon_{a]ce} - A^{0 c}(A_c \cdot A^e) \epsilon_{[ab]e} \nonumber \\
&& + A^0_{[b} A^{0 c} A^{0 e} \epsilon_{a]ce} + A^{0 e} A^{0 c} A^0_e \epsilon_{[a|c|b]}] x_1 \nonumber \\
&& - 2A_{\sigma [a|}A^d_\alpha A^e_\beta \epsilon^{0 \sigma \alpha \beta} \epsilon_{|b]de} (x_3 - x_4) \,,
\end{eqnarray}
which can vanish only if 
\begin{eqnarray}
x_1 &=& 0 \,, \nonumber \\
x_3 - x_4 &=& 0 \,.
\end{eqnarray}
Thus, the Lagrangian that satisfies the constraint algebra is given by
\begin{equation}
\mathcal{L} = x_2 \mathcal{L}_3^2 + x_3 (\mathcal{L}_3^3 + \mathcal{L}_3^4) \,. \label{L14B}
\end{equation}

\subsection{Total derivatives} \label{td14}
Although the Lagrangian in Eq. (\ref{L14B}) satisfies requirements 1 to 4 in Section \ref{requirements}, some of its Lagrangian pieces might be redundant, compared to $\mathcal{L}_2$, via total derivatives.  To find it out, we must proceed to build all the possible total derivatives of currents involving five vector fields. To this end, we must follow a path similar to the ones in previous sections, i.e., employing group theory.  In this way, a term of the form $\partial_\mu (A_\nu A_\rho A_\sigma A_\alpha A_\beta)$, which involves six space-time indices, must be contracted with all the terms in Eqs. (\ref{sixg})-(\ref{sixge}).  However, the Lagrangian pieces we are interested in, $\mathcal{L}_3^2$ and $\mathcal{L}_3^3 + \mathcal{L}_3^4$, explicitly violate parity.  Therefore, only the terms in Eq. (\ref{sixge}) are actually needed.
  This leads to just one term that satisfies the requirement of either being non vanishing or having the potential of being non vanishing once the internal group indices are added:
\begin{equation}
\partial_\mu [(A \cdot A) A_\sigma A_\alpha A_\beta] \epsilon^{\mu \sigma \alpha \beta} \,. \;\;\; (\ast)
\end{equation}
The addition of the internal group indices leads to terms of the form $\partial(A^a A^b A^c A^d A^e)$ that involve five internal group indices.  Therefore, they must be contracted with all the terms in Eq. (\ref{gfive}), which results in
\begin{eqnarray}
\partial_\mu J^\mu_1 &=& \partial_\mu [(A^a \cdot A_a) A_\sigma^c A_\alpha^d A_\beta^e] \epsilon^{\mu \sigma \alpha \beta} \epsilon_{cde} \,, \nonumber \\
\partial_\mu J^\mu_2 &=& \partial_\mu [(A^a \cdot A^b) A_{\sigma a} A_\alpha^d A_\beta^e] \epsilon^{\mu \sigma \alpha \beta} \epsilon_{bde} \,.
\end{eqnarray}
These total derivatives can be expressed in terms of Lagrangian building blocks involving one derivative and four vector fields, which is the key to observe whether some of the two Lagrangian pieces in Eq. (\ref{L14B}) are redundant:
\begin{eqnarray}
\partial_\mu J^\mu_1 &=& \frac{1}{2} [2 A_{\mu \nu}^a A^\nu_a A^c_\sigma A^d_\alpha A^e_\beta \nonumber \\
&& + 3 A_{\mu \sigma}^c A^d_\alpha A^e_\beta (A^a \cdot A_a)] \epsilon^{\mu \sigma \alpha \beta} \epsilon_{cde} \nonumber \\
&& + \mathcal{L}_3^2 \,, \nonumber \\
\partial_\mu J^\mu_2 &=& \frac{1}{2} [A_{\mu \nu}^a A^{\nu b} A_{\sigma a} A^d_\alpha A^e_\beta \nonumber \\
&& +  A_\nu^a A_\mu^{\ \ \nu b} A_{\sigma a} A^d_\alpha A^e_\beta \nonumber \\
&& + (A^a \cdot A^b) A_{\mu \sigma a}  A^d_\alpha A^e_\beta \nonumber \\
&& + 2(A^a \cdot A^b) A_{\sigma a} A_{\mu \alpha}^d  A^e_\beta] \epsilon^{\mu \sigma \alpha \beta} \epsilon_{bde} \nonumber \\
&& + \frac{1}{2} (\mathcal{L}_3^3 + \mathcal{L}_3^4) \,.
\end{eqnarray}
We can see that, even after covariantization, the two Lagrangian pieces in Eq. (\ref{L14B}) can be removed, via total derivatives, in favour of terms already contained in $\mathcal{L}_2$.  Now, from the previous two expressions and the results of Sections \ref{L2} and \ref{Hc14}, we can see that it is legitimate to employ $\partial_\mu J^\mu_1$ and $\partial_\mu J^\mu_2$, since they satisfy the Hessian constraints. Therefore, the conclusion is that there do not exist terms in GSU2P, linearly independent of $\mathcal{L}_2$, that involve one derivative and four vector fields.

\section{Two derivatives} \label{40}

\subsection{The Lagrangian building blocks} \label{building20}
When dealing with two derivatives only, the Lagrangian building blocks, linearly independent of $\mathcal{L}_2$, acquire two possible structures:  either $A_{\mu \nu} S_{\rho \sigma}$ or $S_{\mu \nu} S_{\rho \sigma}$.  In both cases, the number of space-time indices is four, so we have to contract with all the terms in Eq. (\ref{four}).  This results in
\begin{eqnarray}
&&S_\mu^\mu S_\rho^\rho \,, \nonumber \\
&&S_{\mu \nu} S^{\mu \nu} \,,
\end{eqnarray} 
these terms being the only ones that either do not vanish or have the potential of being non vanishing once the internal group indices are added.  Indeed, when this is done, these terms acquire the form $S^a S^b$ which can be contracted only with the group metric $g_{ab}$ \cite{Fuchs:1997jv,Ramond:2010zz,Feger:2012bs}.  Thus, the Lagrangian building blocks are
\begin{eqnarray}
\mathcal{L}_4^1 &=& S_\mu^{\mu a} S_{\rho a}^\rho \,, \nonumber \\
\mathcal{L}_4^2 &=& S_{\mu \nu}^a S^{\mu \nu}_a \,. \label{L202}
\end{eqnarray} 

\subsection{The Hessian constraints} \label{Hc20}
The Lagrangian is therefore written as a linear combination of the Lagrangian building blocks of Eq. (\ref{L202}):
\begin{equation}
\mathcal{L} = \sum_{i=1}^2 x_i \mathcal{L}_4^i \,,
\end{equation}
where the $x_i$ are arbitrary constants.
Since this Lagrangian involves only vector fields through space-time derivatives, the secondary constraint-enforcing relation is satisfied automatically.  Regarding the primary constraint-enforcing relation, the primary Hessian gives the following result:
\begin{equation}
\mathcal{H}^{0 \nu}_{ab} = -8 g_{ab} g^{0 \nu} (x_1 + x_2) \,,
\end{equation}
which vanishes only if
\begin{equation}
x_1 + x_2 = 0 \,.
\end{equation}
Thus, the Lagrangian that satisfies the constraint algebra is given by
\begin{equation}
\mathcal{L} = x_1(\mathcal{L}_4^1 - \mathcal{L}_4^2) \,. \label{L20B}
\end{equation}

\subsection{Total derivatives}
Again, it is absolutely necessary to test if the Lagrangian in Eq. (\ref{L20B}) is not already included in $\mathcal{L}_2$. To this end, it is necessary to build the total derivatives of currents built with one derivative and one vector field.  These terms, being of the form $\partial_\mu [A_\nu (\partial_\rho A_\sigma)]$, involve four space-time indices, so that they are constructed by means of contractions with the terms in Eq.(\ref{four}), except for the last one in that equation as the Lagrangian piece we are interested in, $\mathcal{L}_4^1 - \mathcal{L}_4^2$, explicitly preserves parity.  In this case, none of the terms vanishes, so we end up with three possible total derivatives:
\begin{eqnarray}
&&\partial_\mu [A^\mu (\partial \cdot A)] \,, \nonumber \\
&&\partial_\mu [A_\nu (\partial^\mu A^\nu)] \,, \nonumber \\
&&\partial_\mu [A_\nu (\partial^\nu A^\mu)] \,.
\end{eqnarray}
Since these terms are of the form $\partial[A^a (\partial A^b)]$, once the internal group indices have been added, they can be contracted only with a group metric.  Thus, the total derivatives we have been looking for are
\begin{eqnarray}
\partial_\mu J^\mu_1 &=& \partial_\mu [A^{\mu a} (\partial \cdot A_a)] \,, \nonumber \\
\partial_\mu J^\mu_2 &=& \partial_\mu [A_\nu^a (\partial^\mu A^\nu_a)] \,, \nonumber \\
\partial_\mu J^\mu_3 &=& \partial_\mu [A_\nu^a (\partial^\nu A^\mu_a)] \,.
\end{eqnarray}
It is easy to see that these total derivatives, in their actual form, are anyway useless,
because they lead to terms involving second-order derivatives in addition to the ones we are interested in which involve just two first-order derivatives.  The only way to circumvent this situation, at least partially but enough, is to construct the linear combination
\begin{eqnarray}
\partial_\mu \tilde{J}^\mu_1 &\equiv& \partial_\mu J^\mu_1 - \partial_\mu J^\mu_3 \nonumber \\
&=&-\frac{1}{4} A_{\mu \nu}^a A^{\nu \mu}_a + \frac{1}{4}(\mathcal{L}_4^1 - \mathcal{L}_4^2) +A^{\mu a} [\partial_\mu,\partial_\nu] A^\nu_a \,, \nonumber \\
&& \label{J20}
\end{eqnarray}
that removes the second-order derivatives, since the commutator in the last line trivially vanishes in flat spacetime. Indeed, from this result and the findings in Sections \ref{L2} and \ref{Hc20}, we can see that employing $\partial_\mu \tilde{J}^\mu_1$ is allowed, since it satisfies the Hessian constraints.  The Lagrangian in Eq. (\ref{L20B}) is, in consequence, already contained in $\mathcal{L}_2$ in flat spacetime up to a total derivative. Things, however, are different in curved spacetime.

\subsection{Covariantization} \label{cov}
As is usual the case, the covariantization of Eq. (\ref{J20}) implies the replacement of partial derivatives with space-time covariant derivatives and of the Minkowski metric with an arbitrary space-time metric.  Thus, the curved spacetime version of Eq. (\ref{J20}) reads
\begin{eqnarray}
\nabla_\mu \tilde{J}^\mu_1 &=& -\frac{1}{4} A_{\mu \nu}^a A^{\nu \mu}_a + \frac{1}{4} (\mathcal{L}_4^1 - \mathcal{L}_4^2) +A^{\mu a} [\nabla_\mu,\nabla_\nu] A^\nu_a \nonumber \\
&=& -\frac{1}{4} A_{\mu \nu}^a A^{\nu \mu}_a + \frac{1}{4} (\mathcal{L}_4^1 - \mathcal{L}_4^2) - A^{\mu a} R_{\mu \nu} A^\nu_a \,,
\end{eqnarray}
where $R_{\mu \nu}$ is the Ricci tensor.  Then, we can conclude that the Lagrangian in Eq. (\ref{L20B}) is actually independent of $\mathcal{L}_2$ in a non-redundant way in curved spacetime, whereas it is already included in $\mathcal{L}_2$ in flat spacetime.  To remind the reader of this fact, we will in the following deal with $A^{\mu a} R_{\mu \nu} A^\nu_a$ instead of $\mathcal{L}_4^1 - \mathcal{L}_4^2$.

\subsection{The decoupling limit} \label{declim}
The Helmholtz theorem tells us that any vector field $A_\mu$ can be decomposed into its transverse part, a divergence-free vector field $\mathcal{A}_\mu$, and its longitudinal part, the gradient of scalar field $\nabla_\mu \pi$:
\begin{equation}
A_\mu = \mathcal{A}_\mu + \nabla_\mu \pi \,.
\end{equation}
The decoupling limit of GSU2P, understood as an effective field theory, which corresponds in this case to the replacement $A_\mu^a \rightarrow \nabla_\mu \pi^a$, must also be a healthy theory; i.e., it must be free of the Ostrogradski instability.  Examining the term $A^{\mu a} R_{\mu \nu} A^\nu_a$, we can observe that its decoupling limit $\nabla^\mu \pi^a R_{\mu \nu} \nabla^\nu \pi_a$ is not healthy, as the field equation resultant of the variation of the action with respect to $\pi^a$ leads to a term proportional to $\nabla^\mu R_{\mu \nu}$, i.e., a higher-order term.  To avoid such a pathological behaviour (see Ref. \cite{Deffayet:2009wt}), it is necessary to add $-R(A^a \cdot A_a)/2$ as a counterterm, $R$ being the Ricci scalar:
\begin{empheq}[box=\widefcolorbox]{align}
\mathcal{L}_{4,0} &= A^{\mu a} R_{\mu \nu} A^\nu_a - \frac{1}{2} R(A^a \cdot A_a) \nonumber \\
&= G_{\mu \nu} A^{\mu a} A^\nu_a \,, \label{L20}
\end{empheq}
where $G_{\mu \nu}$ is the Einstein tensor.  Indeed, this Lagrangian is healthy in the decoupling limit because of the divergenceless character of $G_{\mu \nu}$.  Our conclusion, different than the one encountered in Ref. \cite{Allys:2016kbq}, where no term with just two derivatives was found while $G_{\mu \nu} A^{\mu a} A^\nu_a$ was just postulated, finds its origin in the fact that the total derivative in Eq. (\ref{J20}) was first covariantized and later employed (not) to dismiss some terms in favour of others.  This way of proceeding was identified in Ref. \cite{GallegoCadavid:2019zke}, and it is the mechanism to uncover the beyond SU(2) Proca terms as we will later see.  To finish, the notation $\mathcal{L}_{4,0}$ is introduced in Eq. (\ref{L20}) to label this Lagrangian as one that involves (or comes from) two derivatives (this is the reason for the 4) and no vector fields (this is the reason for the 0).

\section{Two derivatives and two vector fields} \label{42}

\subsection{Lagrangian building blocks}
Lagrangian building blocks built from two derivatives and two vector fields are terms of the form $A_{\mu \nu} S_{\rho \sigma} A_\alpha A_\beta$ or $S_{\mu \nu} S_{\rho \sigma} A_\alpha A_\beta$ that involve six space-time indices.  In order to uncover them, we must contract with all the terms in Eqs. (\ref{sixg}) and (\ref{sixge}).  As a result, the Lagrangian building blocks that either do not vanish or have the potential of becoming different than zero once the internal group indices are added are the following:
\begin{eqnarray}
&&A_{\mu \nu} S^\mu_\sigma A^\nu A^\sigma \,, \nonumber \\
&&A_{\mu \nu} S^\rho_\rho A^\mu A^\nu \,, \nonumber \\
&&A_{\mu \nu} S^\nu_\sigma A_\alpha A_\beta \epsilon^{\mu \sigma \alpha \beta} \,, \;\;\; (\ast) \nonumber \\
&&A_{\mu \nu} S^\rho_\rho A_\alpha A_\beta \epsilon^{\mu \nu \alpha \beta} \,, \;\;\; (\ast) \nonumber \\
&&A_{\mu \nu} S_{\rho \sigma} A^\rho A_\beta \epsilon^{\mu \nu \sigma \beta} \,, \nonumber \\
&&S^\mu_\mu S^\rho_\rho (A \cdot A) \,, \nonumber \\
&&S^\mu_\mu S_{\rho \sigma} A^\rho A^\sigma \,, \nonumber \\
&&S_{\mu \nu} S^{\mu \nu} (A \cdot A) \,, \nonumber \\
&&S_{\mu \nu} S^\mu_\sigma A^\nu A^\sigma \,, \nonumber \\
&&S_{\mu \nu} S^\nu_\sigma A_\alpha A_\beta \epsilon^{\mu \sigma \alpha \beta} \,. \;\;\; (\ast)
\end{eqnarray}
When the internal indices are added, these terms are of the form $A^a_{\{\}} S^b A^c A^d$ or $S^a S^b A^c A^d$; i.e., they involve four internal group indices.  So, in order to obtain the Lagrangian building blocks, and according to group theory \cite{Fuchs:1997jv,Ramond:2010zz,Feger:2012bs}, we must contract with the following products of two group metrics:
\begin{eqnarray}
&&g_{ab} g_{cd} \,, \nonumber \\
&&g_{ac} g_{bd} \,, \nonumber \\
&&g_{ad} g_{bc} \,. \label{fourgig}
\end{eqnarray}
This results in the following nineteen Lagrangian building blocks linearly independent of $\mathcal{L}_2$:
\begin{eqnarray}
\mathcal{L}_4^1 &=& A_{\mu \nu}^a S^\mu_{\sigma a} A^{\nu c} A^\sigma_c \,, \nonumber \\
\mathcal{L}_4^2 &=& A_{\mu \nu}^a S^{\mu b}_\sigma A^\nu_a A^\sigma_b \,, \nonumber \\
\mathcal{L}_4^3 &=& A_{\mu \nu}^a S^{\mu b}_\sigma A^\nu_b A^\sigma_a \,, \nonumber \\
\mathcal{L}_4^4 &=& A_{\mu \nu}^a S^{\rho b}_\rho A^\mu_a A^\nu_b \,, \nonumber \\
\mathcal{L}_4^5 &=& A_{\mu \nu}^a S^{\nu b}_\sigma A_{\alpha a} A_{\beta b} \epsilon^{\mu \sigma \alpha \beta} \,, \nonumber \\
\mathcal{L}_4^6 &=& A_{\mu \nu}^a S^{\rho b}_\rho A_{\alpha a} A_{\beta b} \epsilon^{\mu \nu \alpha \beta} \,, \nonumber \\
\mathcal{L}_4^7 &=& A_{\mu \nu}^a S_{\rho \sigma a} A^{\rho c} A_{\beta c} \epsilon^{\mu \nu \sigma \beta} \,, \nonumber \\
\mathcal{L}_4^8 &=& A_{\mu \nu}^a S^b_{\rho \sigma} A^\rho_a A_{\beta b} \epsilon^{\mu \nu \sigma \beta} \,, \nonumber \\
\mathcal{L}_4^9 &=& A_{\mu \nu}^a S^b_{\rho \sigma} A^\rho_b A_{\beta a} \epsilon^{\mu \nu \sigma \beta} \,, \nonumber \\
\mathcal{L}_4^{10} &=& S^{\mu a}_\mu S^\rho_{\rho a} (A^c \cdot A_c) \,, \nonumber \\
\mathcal{L}_4^{11} &=& S^{\mu a}_\mu S^{\rho b}_\rho (A_a \cdot A_b) \,, \nonumber \\
\mathcal{L}_4^{12} &=& S^{\mu a}_\mu S_{\rho \sigma a} A^{\rho c} A^\sigma_c \,, \nonumber \\
\mathcal{L}_4^{13} &=& S^{\mu a}_\mu S^b_{\rho \sigma} A^\rho_a A^\sigma_b \,, \nonumber \\
\mathcal{L}_4^{14} &=& S^a_{\mu \nu} S^{\mu \nu}_a (A^c \cdot A_c) \,, \nonumber \\
\mathcal{L}_4^{15} &=& S^a_{\mu \nu} S^{\mu \nu b} (A_a \cdot A_b) \,, \nonumber \\
\mathcal{L}_4^{16} &=& S^a_{\mu \nu} S^\mu_{\sigma a} A^{\nu c} A^\sigma_c \,, \nonumber \\
\mathcal{L}_4^{17} &=& S^a_{\mu \nu} S^{\mu b}_\sigma A^\nu_a A^\sigma_b \,, \nonumber \\
\mathcal{L}_4^{18} &=& S^a_{\mu \nu} S^{\mu b}_\sigma A^\nu_b A^\sigma_a \,, \nonumber \\
\mathcal{L}_4^{19} &=& S^a_{\mu \nu} S^{\nu b}_\sigma A_{\alpha a} A_{\beta b} \epsilon^{\mu \sigma \alpha \beta} \,.
\end{eqnarray}

\subsection{The Hessian constraints}
The Lagrangian is written as a linear combination of the Lagrangian building blocks found in the previous section.  Thus,
\begin{equation}
\mathcal{L} = \sum_{i = 1}^{19} x_i \mathcal{L}_4^i \,,
\end{equation}
where the $x_i$ are arbitrary constants.  Since the Lagrangian involves two derivatives and two vector fields, none of the Hessian constraints is trivially satisfied in this case.  Performing the calculations, we find for the primary Hessian:
\begin{eqnarray}
\mathcal{H}^{0 \nu}_{ab} &=& -2 A^0_c A^{\nu c} g_{ab} (x_1 + 2 x_{12} + 2 x_{16}) \nonumber \\
&&-2 A^0_c A^{0 c} g^{0\nu} g_{ab} (x_1 - 2 x_{12} - 2 x_{16}) \nonumber \\
&&-2 A^0_a A^\nu_b (x_2 - x_4 + x_{13} + 2 x_{17}) \nonumber \\
&&-2 A^0_a A^0_b g^{0\nu} (x_2 + x_3 - 2 x_{13} - 2 x_{17} - 2 x_{18}) \nonumber \\
&&-2 A^0_b A^\nu_a (x_3 + x_4 + x_{13} + 2 x_{18}) \nonumber \\
&&-2 \epsilon^{0 \nu \alpha \beta} A_{\alpha b} A_{\beta a} (x_5 + 2 x_6 - 2 x_{19}) \nonumber \\
&&-8(A_c \cdot A^c) g^{0\nu} g_{ab} (x_{10} + x_{14}) \nonumber \\
&&-8(A_a \cdot A_b) g^{0\nu} (x_{11} + x_{15}) \,, \label{PH22}
\end{eqnarray}
whereas for the secondary Hessian:
\begin{eqnarray}
\tilde{\mathcal{H}}^{00}_{ab} &=& -2 A^\alpha_{[b|} A^0_{\ \alpha |a]} (x_1 - x_3 - x_4) \nonumber \\
&& - 2 A^\alpha_{[b|} S^0_{\alpha |a]} (2x_{12} - x_{13} + 2 x_{16} - 2 x_{18}) \nonumber \\ 
&& - 2 \epsilon^{\beta \alpha  0 \sigma} A_{\sigma [a|} A_{\beta \alpha |b]} (x_6 - x_7 + x_8) \nonumber \\
&& + 2 A^0_{[a|} S^\alpha_{\alpha |b]} (4 x_{10} - 2 x_{11} + 2 x_{12} - x_{13}) \nonumber \\
&& - 4 A^0_{[a} S^{00}_{b]} (2 x_{14} - x_{15} + x_{16} - x_{17}) \,. \label{SH22}
\end{eqnarray}
Both expressions vanish, therefore, only when the following eleven constraints are satisfied:
\begin{eqnarray}
x_1 &=& 0 \,, \nonumber \\
x_3 &=& - x_2 \,, \nonumber \\
x_4 &=& x_2 \,, \nonumber \\
x_8 &=& - x_6 + x_7 \,, \nonumber \\
x_{13} &=& 4 x_{10} - 2 x_{11} + 2 x_{12} \,, \nonumber \\
x_{14} &=& - x_{10} \,, \nonumber \\
x_{15} &=& - x_{11} \,, \nonumber \\
x_{16} &=& - x_{12} \,, \nonumber \\
x_{17} &=& -2 x_{10} + x_{11} - x_{12} \,, \nonumber \\
x_{18} &=& -2 x_{10} + x_{11} - x_{12} \,, \nonumber \\
x_{19} &=& \frac{x_5}{2} + x_6 \,.
\end{eqnarray}
Thus, the Lagrangian that satisfies the constraint algebra is given by
\begin{eqnarray}
\mathcal{L} &=& x_2 (\mathcal{L}_4^2 - \mathcal{L}_4^3 + \mathcal{L}_4^4) + x_5 \left(\mathcal{L}_4^5 + \frac{\mathcal{L}_4^{19}}{2}\right) \nonumber \\
&& + x_6 (\mathcal{L}_4^6 - \mathcal{L}_4^8 + \mathcal{L}_4^{19}) + x_7 (\mathcal{L}_4^7 + \mathcal{L}_4^8) \nonumber \\
&& + x_9 \mathcal{L}_4^9 + x_{10} (\mathcal{L}_4^{10} + 4 \mathcal{L}_4^{13} - \mathcal{L}_4^{14} - 2 \mathcal{L}_4^{17} - 2 \mathcal{L}_4^{18}) \nonumber \\
&& + x_{11} (\mathcal{L}_4^{11} - 2 \mathcal{L}_4^{13} - \mathcal{L}_4^{15} + \mathcal{L}_4^{17} + \mathcal{L}_4^{18}) \nonumber \\
&& + x_{12} (\mathcal{L}_4^{12} + 2 \mathcal{L}_4^{13} - \mathcal{L}_4^{16} - \mathcal{L}_4^{17} - \mathcal{L}_4^{18}) \,. \label{L22B}
\end{eqnarray}

\subsection{Total derivatives}

With the purpose of establishing which of the Lagrangian pieces in Eq. (\ref{L22B}) are redundant, the total derivatives of terms involving one derivative and three vector fields must be constructed.  These derivatives are terms of the form $\partial_{\mu}[A_\nu (\partial_\rho A_\sigma) A_\alpha A_\beta]$ that involve six space-time indices, so contractions with the terms in Eqs. (\ref{sixg}) and (\ref{sixge}) must be done.  As a result, the only terms that either are different than zero or have the potential of becoming so after introducing the internal group indices are the following:
\begin{eqnarray}
&& \partial_\mu [A^\mu (\partial \cdot A) (A \cdot A)] \,, \nonumber \\
&& \partial_\mu [A^\mu (\partial_\rho A_\sigma) A^\rho A^\sigma] \,, \nonumber \\
&& \partial_\mu [A_\nu (\partial^\mu A^\nu) (A \cdot A)] \,, \nonumber \\
&& \partial_\mu [A_\nu (\partial^\nu A^\mu) (A \cdot A)] \,, \nonumber \\
&& \partial_\mu [A_\nu (\partial^\nu A_\sigma) A_\alpha A_\beta] \epsilon^{\mu \sigma \alpha \beta} \,, \;\;\; (\ast) \nonumber \\
&& \partial_\mu [A_\nu (\partial_\rho A^\nu) A_\alpha A_\beta] \epsilon^{\mu \rho \alpha \beta} \,, \;\;\; (\ast) \nonumber \\
&& \partial_\mu [(\partial_\rho A_\sigma) A_\beta (A \cdot A)] \epsilon^{\mu \rho \sigma \beta} \,, \nonumber \\
&& \partial_\mu [(\partial \cdot A) A_\nu A_\alpha A_\beta] \epsilon^{\mu \nu \alpha \beta} \,. \;\;\; (\ast)
\end{eqnarray}
These total derivatives become terms of the form $\partial[A^a (\partial A^b) A^c A^d]$ once the internal group indices are added.  Since they involve four internal group indices, contractions with the terms in Eq. (\ref{fourgig}) are needed, which results in
\begin{eqnarray}
\partial_\mu J^\mu_1 &=& \partial_\mu [A^{\mu a} (\partial \cdot A_a) (A^c \cdot A_c)] \,, \nonumber \\
\partial_\mu J^\mu_2 &=& \partial_\mu [A^{\mu a} (\partial \cdot A^b) (A_a \cdot A_b)] \,, \nonumber \\
\partial_\mu J^\mu_3 &=& \partial_\mu [A^{\mu a} (\partial_\rho A_{\sigma a}) A^{\rho c} A^\sigma_c] \,, \nonumber \\
\partial_\mu J^\mu_4 &=& \partial_\mu [A^{\mu a} (\partial_\rho A_\sigma^b) A^\rho_a A^\sigma_b] \,, \nonumber \\
\partial_\mu J^\mu_5 &=& \partial_\mu [A^{\mu a} (\partial_\rho A_\sigma^b) A^\rho_b A^\sigma_a] \,, \nonumber \\
\partial_\mu J^\mu_6 &=& \partial_\mu [A_\nu^a (\partial^\mu A^\nu_a) (A^c \cdot A_c)] \,, \nonumber \\
\partial_\mu J^\mu_7 &=& \partial_\mu [A_\nu^a (\partial^\mu A^{\nu b}) (A_a \cdot A_b)] \,, \nonumber \\
\partial_\mu J^\mu_8 &=& \partial_\mu [A_\nu^a (\partial^\nu A^\mu_a) (A^c \cdot A_c)] \,, \nonumber \\
\partial_\mu J^\mu_9 &=& \partial_\mu [A_\nu^a (\partial^\nu A^{\mu b}) (A_a \cdot A_b)] \,, \nonumber \\
\partial_\mu J^\mu_{10} &=& \partial_\mu [A_\nu^a (\partial^\nu A^b_\sigma) A_{\alpha a} A_{\beta b}] \epsilon^{\mu \sigma \alpha \beta} \,, \nonumber \\
\partial_\mu J^\mu_{11} &=& \partial_\mu [A_\nu^a (\partial_\rho A^{\nu b}) A_{\alpha a} A_{\beta b}] \epsilon^{\mu \rho \alpha \beta} \,, \nonumber \\
\partial_\mu J^\mu_{12} &=& \partial_\mu [(\partial_\rho A^a_\sigma) A_{\beta a}  (A^c \cdot A_c)] \epsilon^{\mu \rho \sigma \beta} \,, \nonumber \\
\partial_\mu J^\mu_{13} &=& \partial_\mu [(\partial_\rho A^a_\sigma) A_\beta^b  (A_a \cdot A_b)] \epsilon^{\mu \rho \sigma \beta} \,.
\end{eqnarray}
All these total derivatives are useless as long as they produce terms with second-order derivatives.  Fortunately, this circumstance can be redeemed, although not in all the cases,  by building the following linear combinations:
\begin{eqnarray}
\partial_\mu \tilde{J}^\mu_1 &\equiv& \partial_\mu J^\mu_1 - \partial_\mu J^\mu_8 \nonumber \\
&=& -\frac{1}{4} [A_{\mu \nu}^a A^{\nu \mu}_a (A^c \cdot A_c) + 2 A_\nu^a A^{\nu \mu}_a A_{\mu \rho}^c A^\rho_c] \nonumber \\
&& + \frac{1}{4} (\mathcal{L}_4^{10} - \mathcal{L}_4^{14} + 2 \mathcal{L}_4^{13} - 2\mathcal{L}_4^{17} - 2 \mathcal{L}_4^4) \nonumber \\
&& + A^{\mu a} [\partial_\mu, \partial_\nu] A^\nu_a (A^c \cdot A_c) \,, \nonumber \\
\partial_\mu \tilde{J}^\mu_2 &\equiv& \partial_\mu J^\mu_2 - \partial_\mu J^\mu_9 \nonumber \\
&=& -\frac{1}{4} [A_{\mu \nu}^a A^{\nu \mu b} (A_a \cdot A_b) + A_\nu^a A^{\nu \mu b} A_{\mu \ \  a}^{\ \ \rho} A_{\rho b} \nonumber \\
&& + A_\nu^a A^{\nu \mu b} A_{\mu \rho b} A^\rho_a] \nonumber \\
&& + \frac{1}{4} (\mathcal{L}_4^{11} + \mathcal{L}_4^{13} + \mathcal{L}_4^{12} - \mathcal{L}_4^{15} -  \mathcal{L}_4^{18} -  \mathcal{L}_4^{16} + \mathcal{L}_4^4) \nonumber \\
&& + A^{\mu a} [\partial_\mu, \partial_\nu] A^{\nu b} (A_a \cdot A_b) \,, \nonumber \\
\partial_\mu \tilde{J}^\mu_3 &\equiv& \partial_\mu J^\mu_3 - \partial_\mu J^\mu_5 \nonumber \\
&=& - \frac{1}{4} [A^{\mu a} A_{\rho \sigma}^c A_{\mu \ \  c}^{\ \ \rho} A^\sigma_a + A^{\mu a} A_{\rho \sigma}^c A_{\mu \ \  a}^{\ \ \sigma} A^\rho_c] \nonumber \\
&& + \frac{1}{4} (\mathcal{L}_4^{12} + 2\mathcal{L}_4^{18} - \mathcal{L}_4^{13} - \mathcal{L}_4^{16} \nonumber \\
&&-  \mathcal{L}_4^{17} -  2\mathcal{L}_4^3 - \mathcal{L}_4^4 +  2\mathcal{L}_4^2) \nonumber \\
&& + A^{\mu a} [\partial_\mu, \partial_\rho] A_{\sigma a} A^{\rho c}  A^\sigma_c \,,
\end{eqnarray}
while in the following cases the problem is automatically solved thanks to the symmetries of the Levi-Civita tensor:
\begin{eqnarray}
\partial_\mu \tilde{J}^\mu_4 &\equiv& \partial_\mu J^\mu_{11} = \nonumber \\
&& \frac{1}{4} [A_{\mu \nu}^a A_\rho^{\ \ \nu b} A_{\alpha a} A_{\beta b}  + A_\rho^{\ \ \nu b} A_{\mu \alpha a}  A^a_\nu A_{\beta b} \nonumber \\
&& + A_\rho^{\ \ \nu b} A_{\mu \beta b}  A^a_\nu A_{\alpha a}] \epsilon^{\mu \rho \alpha \beta} \nonumber \\
&& + \frac{1}{4}(\mathcal{L}_4^{19} + 2\mathcal{L}_4^5 - \mathcal{L}_4^8 + \mathcal{L}_4^7) \nonumber \\
&& + \frac{1}{2}A_\nu^a [\partial_\mu, \partial_\rho] A^{\nu b} A_{\alpha a}  A_{\beta b} \epsilon^{\mu \rho \alpha \beta} \,, \nonumber \\
\partial_\mu \tilde{J}^\mu_5 &\equiv& \partial_\mu J^\mu_{12} = \nonumber \\
&& \frac{1}{4} [A_{\rho \sigma}^a A_{\mu \beta a} (A^c \cdot A_c)  + 2 A_{\rho \sigma}^a A_{\mu \alpha}^c A_{\beta a} A^\alpha_c] \epsilon^{\mu \rho \sigma \beta} \nonumber \\
&& + \frac{1}{2} \mathcal{L}_4^9 \nonumber \\
&& + \frac{1}{2} A_{\beta a} [\partial_\mu, \partial_\rho] A_\sigma^a (A^c  \cdot A_c) \epsilon^{\mu \rho \sigma \beta} \,, \nonumber \\
\partial_\mu \tilde{J}^\mu_6 &\equiv& \partial_\mu J^\mu_{13} = \nonumber \\
&& \frac{1}{4} [A_{\rho \sigma}^a A_{\mu \beta}^b (A_a \cdot A_b)  + A_{\rho \sigma}^a A_{\mu \ a}^{\ \alpha} A_\beta^b A_{\alpha b} \nonumber \\
&& + A_{\rho \sigma}^a A_{\mu \alpha b} A_\beta^b A^\alpha_a] \epsilon^{\mu \rho \sigma \beta} \nonumber \\
&& + \frac{1}{4} (\mathcal{L}_4^7 + \mathcal{L}_4^8) \nonumber \\
&& + \frac{1}{2} A_\beta^b [\partial_\mu, \partial_\rho] A_\sigma^a (A_a  \cdot A_b) \epsilon^{\mu \rho \sigma \beta} \,.
\end{eqnarray}
However, even like this, these total derivatives continue to be useless unless they satisfy the Hessian constraints.  Comparison of these expressions with Eqs. (\ref{PH22})-(\ref{SH22}) and with the findings in Section \ref{L2} reveals that the following linear combinations are the only ones that pass the test:
\begin{eqnarray}
&& \partial_\mu(\tilde{J}^\mu_3 + 3 \tilde{J}^\mu_2) \,, \nonumber \\
&& \partial_\mu(2 \tilde{J}^\mu_3 - 3 \tilde{J}^\mu_1) \,, \nonumber \\
&& \partial_\mu \tilde{J}^\mu_5 \,, \nonumber \\
&& \partial_\mu \tilde{J}^\mu_6 \,.  \label{td22}
\end{eqnarray}
We have now the four total derivatives that will help us remove some redundant terms from Eq. (\ref{L22B}).  However, covariantization must be performed first.

\subsection{Covariantization}
The minimal covariantization scheme described in Section \ref{cov} and applied to the total derivatives of Eq. (\ref{td22}) produces the curved space-time versions
\begin{eqnarray}
&&\nabla_\mu(\tilde{J}^\mu_3 + 3 \tilde{J}^\mu_2) = \nonumber \\
&& (... \in \mathcal{L}_2) \nonumber \\
&& + \frac{1}{2} (\mathcal{L}_4^2 - \mathcal{L}_4^3 + \mathcal{L}_4^4) \nonumber \\
&& + \frac{3}{4} (\mathcal{L}_4^{11} - 2\mathcal{L}_4^{13} - \mathcal{L}_4^{15} + \mathcal{L}_4^{17} +  \mathcal{L}_4^{18}) \nonumber \\
&& + (\mathcal{L}_4^{12} + 2\mathcal{L}_4^{13} -  \mathcal{L}_4^{16} - \mathcal{L}_4^{17} - \mathcal{L}_4^{18}) \nonumber \\
&& + A^{\mu a} R^\alpha_{\ \ \sigma \rho \mu} A_{\alpha a} A^{\rho c}  A^\sigma_c - 3 A^{\mu a} R_{\mu \alpha} A^{\alpha b} (A_a \cdot A_b) \,, \nonumber 
\end{eqnarray}
\begin{eqnarray}
&&\nabla_\mu(2 \tilde{J}^\mu_3 - 3 \tilde{J}^\mu_1) = \nonumber \\
&& (... \in \mathcal{L}_2) \nonumber \\
&& + (\mathcal{L}_4^2 - \mathcal{L}_4^3 + \mathcal{L}_4^4) \nonumber \\
&& + \frac{1}{2} (\mathcal{L}_4^{12} + 2\mathcal{L}_4^{13} -  \mathcal{L}_4^{16} - \mathcal{L}_4^{17} - \mathcal{L}_4^{18}) \nonumber \\
&& - \frac{3}{4} (\mathcal{L}_4^{10} + 4\mathcal{L}_4^{13} -  \mathcal{L}_4^{14} - 2 \mathcal{L}_4^{17} - 2 \mathcal{L}_4^{18}) \nonumber \\
&& + 2 A^{\mu a} R^\alpha_{\ \ \sigma \rho \mu} A_{\alpha a} A^{\rho c}  A^\sigma_c + 3 A^{\mu a} R_{\mu \alpha} A^\alpha_a (A^c \cdot A_c) \,, \nonumber 
\end{eqnarray}
\begin{eqnarray}
&&\nabla_\mu \tilde{J}^\mu_5 =  \nonumber \\
&& (... \in \mathcal{L}_2) \nonumber \\
&& + \frac{1}{2} \mathcal{L}_4^9 \nonumber \\
&& + \frac{1}{2} A_{\beta a} R^\alpha_{\ \ \sigma \rho \mu} A_\alpha^a (A^c  \cdot A_c) \epsilon^{\mu \rho \sigma \beta} \,, \nonumber 
\end{eqnarray}
\begin{eqnarray}
&& \nabla_\mu \tilde{J}^\mu_6 =  \nonumber \\
&& (... \in \mathcal{L}_2) \nonumber \\
&& + \frac{1}{4} (\mathcal{L}_4^7 + \mathcal{L}_4^8) \nonumber \\
&& + \frac{1}{2} A_\beta^b R^\alpha_{\ \ \sigma \rho \mu} A_\alpha^a (A_a  \cdot A_b) \epsilon^{\mu \rho \sigma \beta} \,, \label{td42}
\end{eqnarray}
where $(... \in \mathcal{L}_2)$ means terms belonging to $\mathcal{L}_2$ and $R^\alpha_{\ \ \sigma \rho \mu}$ is the Riemann tensor.  We see, therefore, that some terms in Eq. (\ref{L22B}) can be dismissed in flat spacetime but not in curved spacetime.  Indeed, to remind the reader of this difference, these terms will be traded by their respective curvature-dependent companions that appear in the total derivatives in Eq. (\ref{td42}):
\begin{eqnarray}
\mathcal{L} &=& (x_2 + 2 x_{11} - 2x_{12} ) (\mathcal{L}_4^2 - \mathcal{L}_4^3 + \mathcal{L}_4^4) \nonumber \\
&& + \frac{x_5}{2} (2 \mathcal{L}_4^5 + \mathcal{L}_4^{19})  \nonumber \\
&& + x_6 (\mathcal{L}_4^6 - \mathcal{L}_4^8 + \mathcal{L}_4^{19})  \nonumber \\
&& - 2 x_7 A_\beta^b R^\alpha_{\ \ \sigma \rho \mu} A_\alpha^a (A_a  \cdot A_b) \epsilon^{\mu \rho \sigma \beta} \nonumber \\
&& - x_9 A_{\beta a} R^\alpha_{\ \ \sigma \rho \mu} A_\alpha^a (A^c  \cdot A_c) \epsilon^{\mu \rho \sigma \beta} \nonumber \\
&&  + \left(x_{10} - 2x_{11} + \frac{3}{2} x_{12} \right) (\mathcal{L}_4^{10} + 4 \mathcal{L}_4^{13} - \mathcal{L}_4^{14} \nonumber \\
&& - 2 \mathcal{L}_4^{17} - 2 \mathcal{L}_4^{18}) \nonumber \\
&& - \frac{4}{3} x_{11} [A^{\mu a} R^\alpha_{\ \ \sigma \rho \mu} A_{\alpha a} A^{\rho c}  A^\sigma_c \nonumber \\
&& - 3 A^{\mu a} R_{\mu \alpha} A^{\alpha b} (A_a \cdot A_b)] \nonumber \\
&& + \left(\frac{8}{3} x_{11} - 2 x_{12} \right) [2 A^{\mu a} R^\alpha_{\ \ \sigma \rho \mu} A_{\alpha a} A^{\rho c}  A^\sigma_c \nonumber \\
&& + 3 A^{\mu a} R_{\mu \alpha} A^\alpha_a (A^c \cdot A_c)] \,. \label{L42A}
\end{eqnarray}

\subsection{Change of basis} \label{chba}
There are eight linear independent Lagrangian pieces in Eq. (\ref{L42A}) which form a basis set for the construction of the Lagrangian involving two derivatives and two vector fields.  For purposes that will be clear in the following section, we will perform a change of basis that will affect the third and sixth to eighth Lagrangian basis elements in Eq. (\ref{L42A}):
\begin{eqnarray}
\mathcal{L}_4^6 - \mathcal{L}_4^8 + \mathcal{L}_4^{19} && \nonumber \\
&\rightarrow& \mathcal{L}_4^5 - \frac{\mathcal{L}_4^6}{2} + \frac{\mathcal{L}_4^8}{2} \nonumber \\
&& = \frac{1}{2}(2 \mathcal{L}_4^5 + \mathcal{L}_4^{19}) - \frac{1}{2}(\mathcal{L}_4^6 - \mathcal{L}_4^8 + \mathcal{L}_4^{19}) \,, \nonumber 
\end{eqnarray}
\begin{eqnarray}
&& \mathcal{L}_4^{10} + 4 \mathcal{L}_4^{13} - \mathcal{L}_4^{14} - 2 \mathcal{L}_4^{17} - 2 \mathcal{L}_4^{18} \nonumber \\
&& \hspace{5mm} \rightarrow \frac{1}{4}(\mathcal{L}_4^{10} - \mathcal{L}_4^{14} + 2 \mathcal{L}_4^{11} - 2 \mathcal{L}_4^{15}) \nonumber \\
&& \hspace{1cm} = (... \in \mathcal{L}_2) \nonumber \\
&& \hspace{1cm} + (\mathcal{L}_4^2 - \mathcal{L}_4^3 + \mathcal{L}_4^4) \nonumber \\
&& \hspace{1cm} - \frac{3}{4} ( \mathcal{L}_4^{10} + 4 \mathcal{L}_4^{13} - \mathcal{L}_4^{14} - 2 \mathcal{L}_4^{17} - 2 \mathcal{L}_4^{18}) \nonumber \\
&& \hspace{1cm} + \frac{2}{3} \Big\{ \nabla_\mu(\tilde{J}^\mu_3 + 3 \tilde{J}^\mu_2) - [A^{\mu a} R^\alpha_{\ \ \sigma \rho \mu} A_{\alpha a} A^{\rho c}  A^\sigma_c \nonumber \\
&& \hspace{1cm} - 3 A^{\mu a} R_{\mu \alpha} A^{\alpha b} (A_a \cdot A_b)] \Big\} \nonumber \\
&& \hspace{1cm} - \frac{4}{3} \Big\{ \nabla_\mu(2\tilde{J}^\mu_3 - 3 \tilde{J}^\mu_1) - [2 A^{\mu a} R^\alpha_{\ \ \sigma \rho \mu} A_{\alpha a} A^{\rho c}  A^\sigma_c \nonumber \\
&& \hspace{1cm} + 3 A^{\mu a} R_{\mu \alpha} A^\alpha_a (A^c \cdot A_c)] \Big\} \,, \nonumber
\end{eqnarray}
\begin{eqnarray}
&& A^{\mu a} R^\alpha_{\ \ \sigma \rho \mu} A_{\alpha a} A^{\rho c}  A^\sigma_c - 3 A^{\mu a} R_{\mu \alpha} A^{\alpha b} (A_a \cdot A_b) \nonumber \\
&& \hspace{5mm} \rightarrow A^{\mu a} R^\alpha_{\ \ \sigma \rho \mu} A_{\alpha a} A^{\rho c}  A^\sigma_c - 3 A^{\mu a} R_{\mu \alpha} A^{\alpha b} (A_a \cdot A_b) \nonumber \\
&& \hspace{1cm} + \tilde{a} \left[ \frac{1}{4}(\mathcal{L}_4^{10} - \mathcal{L}_4^{14} + 2 \mathcal{L}_4^{11} - 2 \mathcal{L}_4^{15}) \right] \,, \nonumber
\end{eqnarray}
\begin{eqnarray}
&& 2 A^{\mu a} R^\alpha_{\ \ \sigma \rho \mu} A_{\alpha a} A^{\rho c}  A^\sigma_c  + 3 A^{\mu a} R_{\mu \alpha} A^\alpha_a (A^c \cdot A_c) \nonumber \\
&& \hspace{5mm} \rightarrow 2 A^{\mu a} R^\alpha_{\ \ \sigma \rho \mu} A_{\alpha a} A^{\rho c}  A^\sigma_c  + 3 A^{\mu a} R_{\mu \alpha} A^\alpha_a (A^c \cdot A_c) \nonumber \\
&& \hspace{1cm} + \tilde{b} \left[ \frac{1}{4}(\mathcal{L}_4^{10} - \mathcal{L}_4^{14} + 2 \mathcal{L}_4^{11} - 2 \mathcal{L}_4^{15}) \right] \,,
\end{eqnarray}
where $\tilde{a}$ and $\tilde{b}$ are arbitrary constants.
Thus, the Lagrangian involving two derivatives and two vector fields is written as follows:
\begin{equation}
\mathcal{L} = \sum_{i = 1}^8 \hat{\alpha}_i \hat{\mathcal{L}}_4^i \,, \label{L42WDL}
\end{equation}
with
\begin{eqnarray}
\hat{\mathcal{L}}_4^1 &=& \frac{1}{4}(\mathcal{L}_4^{10} - \mathcal{L}_4^{14} + 2 \mathcal{L}_4^{11} - 2 \mathcal{L}_4^{15}) \,, \nonumber \\
\hat{\mathcal{L}}_4^2 &=& \mathcal{L}_4^2 - \mathcal{L}_4^3 + \mathcal{L}_4^4 \,, \nonumber \\
\hat{\mathcal{L}}_4^3 &=& A^{\mu a} R^\alpha_{\ \ \sigma \rho \mu} A_{\alpha a} A^{\rho c}  A^\sigma_c - 3 A^{\mu a} R_{\mu \alpha} A^{\alpha b} (A_a \cdot A_b) \nonumber \\
&& + \tilde{a} \left[ \frac{1}{4}(\mathcal{L}_4^{10} - \mathcal{L}_4^{14} + 2 \mathcal{L}_4^{11} - 2 \mathcal{L}_4^{15}) \right] \,, \nonumber \\
\hat{\mathcal{L}}_4^4 &=& 2 A^{\mu a} R^\alpha_{\ \ \sigma \rho \mu} A_{\alpha a} A^{\rho c}  A^\sigma_c + 3 A^{\mu a} R_{\mu \alpha} A^\alpha_a (A^c \cdot A_c) \nonumber \\
&& + \tilde{b} \left[ \frac{1}{4}(\mathcal{L}_4^{10} - \mathcal{L}_4^{14} + 2 \mathcal{L}_4^{11} - 2 \mathcal{L}_4^{15}) \right] \,, \nonumber \\
\hat{\mathcal{L}}_4^5 &=& 2 \mathcal{L}_4^5 + \mathcal{L}_4^{19} \,, \nonumber \\
\hat{\mathcal{L}}_4^6 &=& \mathcal{L}_4^5 - \frac{\mathcal{L}_4^6}{2} + \frac{\mathcal{L}_4^8}{2} \,, \nonumber \\
\hat{\mathcal{L}}_4^7 &=& A_\beta^b R^\alpha_{\ \ \sigma \rho \mu} A_\alpha^a (A_a  \cdot A_b) \epsilon^{\mu \rho \sigma \beta} \,, \nonumber \\
\hat{\mathcal{L}}_4^8 &=& A_{\beta a} R^\alpha_{\ \ \sigma \rho \mu} A_\alpha^a (A^c  \cdot A_c) \epsilon^{\mu \rho \sigma \beta} \,, \label{L42WDLC}
\end{eqnarray}
where the $\hat{\alpha}_i$ are arbitrary constants.  We have deliberately ordered the Lagrangian pieces this way so that the first four are the ones that preserve parity while the last four, in contrast, are the ones that do not preserve it.

\subsection{The decoupling limit}

Following the general description of Section \ref{declim}, the decoupling limit of the theory described by Eqs. (\ref{L42WDL}) and (\ref{L42WDLC}), obtained by making the replacement $A_\mu^a \rightarrow \nabla_\mu \pi^a$, must be free of the Ostrogradski instability.  This is easy to verify for $\hat{\mathcal{L}}_4^2$ and $\hat{\mathcal{L}}_4^6$ whose decoupling limits vanish thanks to the antisymmetry of $A_{\mu \nu}^a$.  It is also easy to verify for $\hat{\mathcal{L}}_4^7$ and $\hat{\mathcal{L}}_4^8$ having in mind their relation to $\nabla_\mu \tilde{J}^\mu_6$ and $\nabla_\mu \tilde{J}^\mu_5$, respectively, as shown in Eq. (\ref{td42}), and, again, the antisymmetry of $A_{\mu \nu}^a$.  Now, regarding $\hat{\mathcal{L}}_4^1$, its decoupling limit leads to higher-order field equations, because, contrary to partial derivatives, covariant derivatives do not commute.  This can be redeemed by adding a specific counterterm so that the healthy version of $\hat{\mathcal{L}}_4^1$ becomes:
\begin{eqnarray}
\hat{\mathcal{L}}_4^{1,h} &=& \frac{1}{4} (A_b \cdot A^b) [S^{\mu a}_\mu S^\nu_{\nu a} - S^{\mu a}_\nu S^\nu_{\mu a} - R (A^a \cdot A_a)] \nonumber \\
&& + \frac{1}{2} (A_a \cdot A_b) [S^{\mu a}_\mu S^{\nu b}_\nu - S^{\mu a}_\nu S^{\nu b}_\mu - R (A^a \cdot A^b)] \,. \nonumber \\
&& \label{L41h}
\end{eqnarray}
In contrast, although the decoupling limit of $\hat{\mathcal{L}}_4^5$, specifically the term $\mathcal{L}_4^{19}$, leads as well to higher-order field equations, it turned out impossible to find out the required counterterm\footnote{The isolation of $\mathcal{L}_4^{19}$ in just one Lagrangian piece is motivated by the impossibility of finding out a counterterm, and it is the reason of the first change in basis elements shown in the previous section.}.  This leaves us with two possibilities:  either we must discard $\hat{\mathcal{L}}_4^5$, as it is pathological in the decoupling limit, or we must keep it, because its decoupling limit is degenerate and this property might, in principle, remove the ghostly degree of freedom \cite{Langlois:2015cwa,Ganz:2020skf}.  We will not know which possibility is the right one until a proper and dedicated analysis of the degeneracy conditions in the decoupling limit is performed\footnote{This seems quite non trivial, so we rather leave it for future work.}.  Finally, $\hat{\mathcal{L}}_4^3$ and $\hat{\mathcal{L}}_4^4$ are the non-Abelian versions of a term in the generalized Proca theory identified unequivocally in Ref. \cite{GallegoCadavid:2019zke} as the beyond Proca term \cite{Heisenberg:2016eld}.  We conjecture then that $\hat{\mathcal{L}}_4^3$ and $\hat{\mathcal{L}}_4^4$ are the beyond generalized SU(2) Proca terms whose decoupling limits must satisfy all the conditions required to remove the Ostrogradski ghosts.  This fixes the $\tilde{a}$ and $\tilde{b}$ constants, but, since the non-Abelian extension of the beyond multi-Galileon theory has not been constructed yet, the actual values of $\tilde{a}$ and $\tilde{b}$ are unknown to us.  To circumvent this lack of knowledge, we can take advantage of the fact that, although the Abelian and non-Abelian vector-tensor theories are different\footnote{Abelian theories display some terms whose non-Abelian versions do not exist and vice versa.} despite sharing many aspects in their construction, the non-Abelian theory stripped of the internal group indices must be contained in the Abelian theory.  Thus, once $\hat{\mathcal{L}}_4^3$ is stripped of the internal group indices, it becomes
\begin{equation}
\hat{\mathcal{L}}_4^3 \rightarrow  - 3 A^\mu R_{\mu \alpha} A^\alpha A^2 + \tilde{a} \frac{3}{4} A^2 (S^\mu_\mu S^\nu_\nu - S^\mu_\nu S^\nu_\mu) \,,
\end{equation}
which must be compared with Eq. (42) in Ref. \cite{GallegoCadavid:2019zke}:\footnote{This is the reason of the third and fourth changes in basis elements shown in the previous section.}
\begin{eqnarray}
\mathcal{L}_4^{BP} &=& G_N(X) R_{\mu \nu} A^\mu A^\nu \nonumber \\
&& - [2X G_{N,X} (X) + G_N(X)] \frac{1}{4} (S^\mu_\mu S^\nu_\nu - S^\mu_\nu S^\nu_\mu) \,, \nonumber \\
&& \label{GBP}
\end{eqnarray}
where $X = -A^2/2$, $G_N(X)$ is an arbitrary function of $X$, and $G_{N,X} (X)$ is the derivative of $G_N(X)$ with respect to $X$.  We see that these two Lagrangian pieces are equivalent for $G_N(X) = 6 X$ and $\tilde{a} = 3$.  Similarly, once $\hat{\mathcal{L}}_4^4$ is stripped of the internal group indices, it becomes
\begin{equation}
\hat{\mathcal{L}}_4^4 \rightarrow  3 A^\mu R_{\mu \alpha} A^\alpha A^2 + \tilde{b} \frac{3}{4} A^2 (S^\mu_\mu S^\nu_\nu - S^\mu_\nu S^\nu_\mu) \,,
\end{equation}
which is equivalent to the Lagrangian piece in Eq. (\ref{GBP}) for $G_N(X) = - 6 X$ and $\tilde{b} = - 3$.

\subsection{A new change of basis}

Having found the actual values for $\tilde{a}$ and $\tilde{b}$ in the previous section, $\hat{\mathcal{L}}_4^3$ and $\hat{\mathcal{L}}_4^4$ acquire the form
\begin{eqnarray}
\hat{\mathcal{L}}_4^3 &=& A^{\mu a} R^\alpha_{\ \ \sigma \rho \mu} A_{\alpha a} A^{\rho c}  A^\sigma_c - 3 A^{\mu a} R_{\mu \alpha} A^{\alpha b} (A_a \cdot A_b) \nonumber \\
&& + 3 \Big[\frac{1}{4} (A_b \cdot A^b) (S^{\mu a}_\mu S^\nu_{\nu a} - S^{\mu a}_\nu S^\nu_{\mu a}) \nonumber \\
&& + \frac{1}{2} (A_a \cdot A_b) (S^{\mu a}_\mu S^{\nu b}_\nu - S^{\mu a}_\nu S^{\nu b}_\mu) \Big] \,, \nonumber \\
\hat{\mathcal{L}}_4^4 &=& 2 A^{\mu a} R^\alpha_{\ \ \sigma \rho \mu} A_{\alpha a} A^{\rho c}  A^\sigma_c + 3 A^{\mu a} R_{\mu \alpha} A^\alpha_a (A^c \cdot A_c) \nonumber \\
&& - 3 \Big[\frac{1}{4} (A_b \cdot A^b) (S^{\mu a}_\mu S^\nu_{\nu a} - S^{\mu a}_\nu S^\nu_{\mu a}) \nonumber \\
&& + \frac{1}{2} (A_a \cdot A_b) (S^{\mu a}_\mu S^{\nu b}_\nu - S^{\mu a}_\nu S^{\nu b}_\mu) \Big] \,,
\end{eqnarray}
which can be replaced by
\begin{eqnarray}
\hat{\mathcal{L}}_4^3 && \nonumber \\
&\rightarrow& A^{\mu a} R^\alpha_{\ \ \sigma \rho \mu} A_{\alpha a} A^{\rho c}  A^\sigma_c + \frac{3}{4} (A_b \cdot A^b) (A^a \cdot A_a) R \nonumber \\
&& = \hat{\mathcal{L}}_4^3 - 3 \hat{\mathcal{L}}_4^{1,h} + 3 G_{\mu \nu} A^{\mu a} A^{\nu b} (A_a \cdot A_b) \,, \nonumber \\
\hat{\mathcal{L}}_4^4 && \nonumber \\
&\rightarrow& 2 A^{\mu a} R^\alpha_{\ \ \sigma \rho \mu} A_{\alpha a} A^{\rho c}  A^\sigma_c  \nonumber \\
&& + \frac{3}{4} [(A_b \cdot A^b) (A^a \cdot A_a) - 2 (A_a \cdot A_b) (A^a \cdot A^b)] R \nonumber \\
&& = \hat{\mathcal{L}}_4^4 + 3 \hat{\mathcal{L}}_4^{1,h} - 3 G_{\mu \nu} A^{\mu a} A^\nu_a (A^b \cdot A_b) \,,
\end{eqnarray}
where we have added and subtracted, respectively, the Lagrangian pieces $G_{\mu \nu} A^{\mu a} A^{\nu b} (A_a \cdot A_b)$ and $G_{\mu \nu} A^{\mu a} A^\nu_a (A^b \cdot A_b)$ that exist only in curved spacetime and whose decoupling limit is healthy, since $G_{\mu \nu}$ is divergenceless.  Furthermore, we can replace the second Lagrangian piece in the previous expression as follows:
\begin{eqnarray}
&&\hat{\mathcal{L}}_4^4 + 3 \hat{\mathcal{L}}_4^{1,h} - 3 G_{\mu \nu} A^{\mu a} A^\nu_a (A^b \cdot A_b) \nonumber \\
&& \hspace{5mm} \rightarrow - \frac{3}{4} [(A_b \cdot A^b) (A^a \cdot A_a) + 2 (A_a \cdot A_b) (A^a \cdot A^b)] R \nonumber \\
&&  \hspace{1cm} = \hat{\mathcal{L}}_4^4 + 3 \hat{\mathcal{L}}_4^{1,h} - 3 G_{\mu \nu} A^{\mu a} A^\nu_a (A^b \cdot A_b) \nonumber \\
&& \hspace{1cm} - 2 [\hat{\mathcal{L}}_4^3 - 3 \hat{\mathcal{L}}_4^{1,h} + 3 G_{\mu \nu} A^{\mu a} A^{\nu b} (A_a \cdot A_b)] \,,
\end{eqnarray}
which is indeed very interesting, because now $\hat{\mathcal{L}}_4^{1,h}$ can be replaced by
\begin{eqnarray}
&& \hat{\mathcal{L}}_4^{1,h} \nonumber \\
&& \hspace{2mm} \rightarrow \frac{1}{4} \Big\{(A_b \cdot A^b) [S^{\mu a}_\mu S^\nu_{\nu a} - S^{\mu a}_\nu S^\nu_{\mu a}] \nonumber \\
&& \hspace{7mm} + 2 (A_a \cdot A_b) [S^{\mu a}_\mu S^{\nu b}_\nu - S^{\mu a}_\nu S^{\nu b}_\mu] \Big\} \nonumber \\
&& \hspace{7mm} = 3 \hat{\mathcal{L}}_4^{1,h} + \frac{3}{4} [(A_b \cdot A^b) (A^a \cdot A_a) \nonumber \\
&& \hspace{7mm} + 2 (A_a \cdot A_b) (A^a \cdot A^b)] R \,,
\end{eqnarray}
this just being the original $\mathcal{L}_4^1$, i.e., without its respective counterterm.

All together, we can formulate the reconstructed GSU2P Lagrangian composed of two derivatives and two vector fields as follows:
\begin{empheq}[box=\widefcolorbox]{align}
\mathcal{L}_{4,2} = \sum_{i = 1}^6 \frac{\alpha_i}{m_P^2} \mathcal{L}_{4,2}^i + \sum_{i = 1}^4 \frac{\tilde{\alpha}_i}{m_P^2} \tilde{\mathcal{L}}_{4,2}^i \,, \label{L42}
\end{empheq}
where
\begin{empheq}[box=\widefcolorbox]{align}
\mathcal{L}_{4,2}^1 =& (A_b \cdot A^b) [S^{\mu a}_\mu S^\nu_{\nu a} - S^{\mu a}_\nu S^\nu_{\mu a}] \nonumber \\
& + 2 (A_a \cdot A_b) [S^{\mu a}_\mu S^{\nu b}_\nu - S^{\mu a}_\nu S^{\nu b}_\mu] \,, \nonumber \\
\mathcal{L}_{4,2}^2 =& A_{\mu \nu}^a S^{\mu b}_\sigma A^\nu_a A^\sigma_b - A_{\mu \nu}^a S^{\mu b}_\sigma A^\nu_b A^\sigma_a + A_{\mu \nu}^a S^{\rho b}_\rho A^\mu_a A^\nu_b \,, \nonumber \\
\mathcal{L}_{4,2}^3 =& A^{\mu a} R^\alpha_{\ \ \sigma \rho \mu} A_{\alpha a} A^{\rho b}  A^\sigma_b + \frac{3}{4} (A_b \cdot A^b) (A^a \cdot A_a) R \,, \nonumber \\
\mathcal{L}_{4,2}^4 =& [(A_b \cdot A^b) (A^a \cdot A_a) + 2 (A_a \cdot A_b) (A^a \cdot A^b)] R \,, \nonumber \\
\mathcal{L}_{4,2}^5 =& G_{\mu \nu} A^{\mu a} A^\nu_a (A^b \cdot A_b) \,, \nonumber \\
\mathcal{L}_{4,2}^6 =& G_{\mu \nu} A^{\mu a} A^{\nu b} (A_a \cdot A_b) \,, \label{L42LPa}
\end{empheq}
\begin{empheq}[box=\widefcolorbox]{align}
\tilde{\mathcal{L}}_{4,2}^1 =& -2 A_{\mu \nu}^a S^{\mu b}_\sigma A_{\alpha a} A_{\beta b} \epsilon^{\nu \sigma \alpha \beta} + S^a_{\mu \nu} S^{\nu b}_\sigma A_{\alpha a} A_{\beta b} \epsilon^{\mu \sigma \alpha \beta} \,, \nonumber \\
\tilde{\mathcal{L}}_{4,2}^2 =& A_{\mu \nu}^a S^{\mu b}_\sigma A_{\alpha a} A_{\beta b} \epsilon^{\nu \sigma \alpha \beta} - \tilde{A}^{\alpha \beta}_a S^b_{\rho \alpha} A^{\rho a} A_{\beta b} \nonumber \\
& + \tilde{A}^{\alpha \beta}_a S^\rho_{\rho b} A_\alpha^a A_\beta^b \,, \nonumber \\
\tilde{\mathcal{L}}_{4,2}^3 =& A_\beta^b R^\alpha_{\ \ \sigma \rho \mu} A_\alpha^a (A_a  \cdot A_b) \epsilon^{\mu \rho \sigma \beta} \,, \nonumber \\
\tilde{\mathcal{L}}_{4,2}^4 =& A_{\beta a} R^\alpha_{\ \ \sigma \rho \mu} A_\alpha^a (A^b  \cdot A_b) \epsilon^{\mu \rho \sigma \beta} \,, \label{L42LPb}
\end{empheq}
the $\alpha_i$ and $\tilde{\alpha}_i$ being arbitrary dimensionless constants, $m_P$ being the reduced Planck mass, $\tilde{A}^{\mu \nu}_a \equiv \frac{1}{2} \epsilon^{\mu \nu \rho \sigma} A_{\rho \sigma a}$ being the Hodge dual of $A_{\mu \nu}^a$, and the Lagrangian pieces having been deliberately split into those that preserve parity (the ones without a tilde) and those that do not preserve it (the ones with a tilde).  It is worthwhile mentioning that the subscripts 4,2 have been introduced to remind the reader that two derivatives and two vector fields have been employed to build the different Lagrangian pieces.

\section{Three derivatives} \label{50}

\subsection{Lagrangian building blocks}

Terms of the form $A_{\mu \nu} A_{\rho \sigma} S_{\alpha \beta}$, $A_{\mu \nu} S_{\rho \sigma} S_{\alpha \beta}$, and $S_{\mu \nu} S_{\rho \sigma} S_{\alpha \beta}$, that involve six space-time indices, are the ones that become the Lagrangian building blocks of a Lagrangian built with just three derivatives once they are contracted with the terms in Eqs. (\ref{sixg}) and (\ref{sixge}).  Upon the contractions, the only blocks that either do not vanish or have the potential of becoming non vanishing once the internal group indices are introduced are the following:
\begin{eqnarray}
&& A_{\mu \nu} A^{\mu \nu} S^\alpha_\alpha \,, \nonumber \\
&& A_{\mu \nu} A^\mu_{\ \ \sigma} S^{\nu \sigma} \,, \nonumber \\
&& A_{\mu \nu} A_{\rho \sigma} S^\nu_\beta \epsilon^{\mu \rho \sigma \beta} \,, \nonumber \\
&& A_{\mu \nu} A_{\rho \sigma} S^\alpha_\alpha \epsilon^{\mu \nu \rho \sigma} \,, \nonumber \\
&& A_{\mu \nu} S^\mu_\sigma S^{\nu \sigma} \,, \;\;\; (\ast) \nonumber \\
&& A_{\mu \nu} S_{\rho \sigma} S^\rho_\beta \epsilon^{\mu \nu \sigma \beta} \,, \;\;\; (\ast) \nonumber \\
&& S^\mu_\mu S^\rho_\rho S^\alpha_\alpha \,, \nonumber \\
&& S^\mu_\mu S_{\rho \sigma} S^{\rho \sigma} \,, \nonumber \\
&& S^{\mu \nu} S^\mu_\sigma S^{\nu \sigma} \,.
\end{eqnarray}
The introduction of the internal group indices makes these terms become of the form $A^a_{\{\}} A^b_{\{\}} S^c$, $A^a_{\{\}} S^b S^c$, or $S^a S^b S^c$, involving three internal group indices, which lead to group-invariant Lagrangian building blocks upon contractions with $\epsilon_{abc}$.  Most of these blocks, however, vanish because of the antisymmetric nature of $\epsilon_{abc}$, the only survivals being
\begin{eqnarray}
\mathcal{L}_5^1 &=& A_{\mu \nu}^a A_{\rho \sigma}^b S^{\nu c}_\beta \epsilon^{\mu \rho \sigma \beta} \epsilon_{abc} \,, \nonumber \\
\mathcal{L}_5^2 &=& A_{\mu \nu}^a S^{\mu b}_\sigma S^{\nu \sigma c} \epsilon_{abc} \,, \nonumber \\
\mathcal{L}_5^3 &=& A_{\mu \nu}^a S_{\rho \sigma}^b S^{\rho c}_\beta \epsilon^{\mu \nu \sigma \beta} \epsilon_{abc} \,. \label{L30}
\end{eqnarray}

\subsection{The Hessian constraints} \label{Hc30}
The linear combination
\begin{equation}
\mathcal{L} = \sum_{i = 1}^3 x_i \mathcal{L}_5^i \,,
\end{equation}
where the $x_i$ are arbitrary constants and the $\mathcal{L}_5^i$ are the ones in Eq. (\ref{L30}), makes the GSU2P Lagrangian built with just three derivatives.  Because no single vector field appears in this Lagrangian, the secondary constraint-enforcing relation is trivially satisfied.  Regarding the primary constraint-enforcing relation, the primary Hessian gives the following result:
\begin{eqnarray}
\mathcal{H}^{0 \nu}_{ab} &=& 2 A_{\rho \sigma}^c \epsilon^{\nu \rho \sigma 0} \epsilon_{bca} (x_1 + 2 x_3) \nonumber \\
&& + 4 (S^{0 \nu c} + g^{0 \nu} S^{00 c} - A^{0 \nu c}) \epsilon_{abc} x_2 \,,
\end{eqnarray}
which vanishes only if 
\begin{eqnarray}
x_1 + 2 x_3 &=& 0 \,, \nonumber \\
x_2 &=& 0 \,.
\end{eqnarray}
The Lagrangian that satisfies the constraint algebra is, therefore,
\begin{equation}
\mathcal{L} = x_3 (- 2 \mathcal{L}_5^1 + \mathcal{L}_5^3) \,. \label{L30B}
\end{equation}

\subsection{Total derivatives}

As with the other Lagrangians involving a different number of derivatives and/or vector fields, we must be sure that the Lagrangian in Eq. (\ref{L30B}) is not redundant compared with terms in $\mathcal{L}_2$.  To this end, we must construct total derivatives of terms involving two derivatives and one vector field, i.e., total derivatives of the form $\partial_\mu[A_\nu (\partial_\rho A_\sigma) (\partial_\alpha A_\beta)]$.  These terms involve six space-time indices, so that they must be contracted with those terms in Eq. (\ref{sixg}) and (\ref{sixge}).  However, since the Lagrangian in Eq. (\ref{L30B}) does not preserve parity, it will be enough to contract with the terms in Eq. (\ref{sixge}).  Thus, the only terms that either are non vanishing or can become non vanishing once the internal group indices are added are the following:
\begin{eqnarray}
&& \partial_\mu[A_\nu (\partial^\nu A_\sigma) (\partial_\alpha A_\beta)] \epsilon^{\mu \sigma \alpha \beta} \,, \nonumber \\
&& \partial_\mu[A_\nu (\partial_\rho A^\nu) (\partial_\alpha A_\beta)] \epsilon^{\mu \rho \alpha \beta} \,, \nonumber \\
&& \partial_\mu[A_\nu (\partial \cdot A) (\partial_\alpha A_\beta)] \epsilon^{\mu \nu \alpha \beta} \,, \nonumber \\
&& \partial_\mu[A_\nu (\partial_\rho A_\sigma) (\partial^\rho A_\beta)] \epsilon^{\mu \nu \sigma \beta} \,, \;\;\; (\ast) \nonumber \\
&& \partial_\mu[A_\nu (\partial_\rho A_\sigma) (\partial_\alpha A^\rho)] \epsilon^{\mu \nu \sigma \alpha} \,, \nonumber \\
&& \partial_\mu[A_\nu (\partial_\rho A_\sigma) (\partial_\alpha A^\sigma)] \epsilon^{\mu \nu \rho \alpha} \,, \;\;\; (\ast)
\end{eqnarray}
which, in turn, can be contracted only with $\epsilon_{abc}$ after adding the internal group indices, since the total derivatives acquire the form $\partial[A^a(\partial A^b) (\partial A^c)]$:
\begin{eqnarray}
\partial_\mu J^\mu_1 &=& \partial_\mu[A_\nu^a (\partial^\nu A_\sigma^b) (\partial_\alpha A_\beta^c)] \epsilon^{\mu \sigma \alpha \beta} \epsilon_{abc} \,, \nonumber \\
\partial_\mu J^\mu_2 &=& \partial_\mu[A_\nu^a (\partial_\rho A^{\nu b}) (\partial_\alpha A_\beta^c)] \epsilon^{\mu \rho \alpha \beta} \epsilon_{abc} \,, \nonumber \\
\partial_\mu J^\mu_3 &=& \partial_\mu[A_\nu^a (\partial \cdot A^b) (\partial_\alpha A_\beta^c)] \epsilon^{\mu \nu \alpha \beta} \epsilon_{abc} \,, \nonumber \\
\partial_\mu J^\mu_4 &=& \partial_\mu[A_\nu^a (\partial_\rho A_\sigma^b) (\partial^\rho A_\beta^c)] \epsilon^{\mu \nu \sigma \beta} \epsilon_{abc} \,, \nonumber \\
\partial_\mu J^\mu_5 &=& \partial_\mu[A_\nu^a (\partial_\rho A_\sigma^b) (\partial_\alpha A^{\rho c})] \epsilon^{\mu \nu \sigma \alpha} \epsilon_{abc} \,, \nonumber \\
\partial_\mu J^\mu_6 &=& \partial_\mu[A_\nu^a (\partial_\rho A_\sigma^b) (\partial_\alpha A^{\sigma c})] \epsilon^{\mu \nu \rho \alpha} \epsilon_{abc} \,. 
\end{eqnarray}
As the reader has already learned, these total derivatives are completely useless unless the second derivatives they produce may be canceled out.  After a careful observation of these terms, only two are able by themselves to get rid of the second derivatives in flat spacetime thanks to the antisymmetry of the Levi-Civita tensor:
\begin{eqnarray}
\partial_\mu \tilde{J}^\mu_1 &\equiv& \partial_\mu J^\mu_2 = \nonumber \\
&& \frac{1}{8} A_{\mu \nu}^a A_\rho^{\ \ \nu b} A_{\alpha \beta}^c \epsilon^{\mu \rho \alpha \beta} \epsilon_{abc} \nonumber \\
&& +  \frac{1}{8} (- 2 \mathcal{L}_5^1 + \mathcal{L}_5^3) \nonumber \\
&& + \frac{1}{4} \{ A_\nu^a [\partial_\mu,\partial_\rho] A^{\nu b} A_{\alpha \beta}^c + A_\nu^a A_\rho^{\ \ \nu b} [\partial_\mu,\partial_\alpha] A_\beta^c \nonumber \\
&& + A_\nu^a S_\rho^{\nu b} [\partial_\mu,\partial_\alpha] A_\beta^c \} \epsilon^{\mu \rho \alpha \beta} \epsilon_{abc} \,, \nonumber \\
\partial_\mu \tilde{J}^\mu_2 &\equiv& \partial_\mu J^\mu_6 = \nonumber \\
&& \frac{1}{8} A_{\mu \nu}^a A_{\rho \sigma}^b A_\alpha^{\ \ \sigma c} \epsilon^{\mu \nu \rho \alpha} \epsilon_{abc} \nonumber \\
&& +  \frac{1}{8} (- 2 \mathcal{L}_5^1 + \mathcal{L}_5^3) \nonumber \\
&& + \frac{1}{2} \{ A_\nu^a [\partial_\mu,\partial_\rho] A_\sigma^b A_\alpha^{\ \ \sigma c} \nonumber \\
&& + A_\nu^a [\partial_\mu,\partial_\rho] A_\sigma^b S_\alpha^{\sigma c} \} \epsilon^{\mu \nu \rho \alpha} \epsilon_{abc} \,. \label{td30}
\end{eqnarray}
Indeed, from this result and the findings in Sections \ref{L2} and \ref{Hc30}, we can see that employing either $\partial_\mu \tilde{J}^\mu_1$ or $\partial_\mu \tilde{J}^\mu_2$ is allowed, since they satisfy the Hessian constraints.  The conclusion is that the Lagrangian in Eq. (\ref{L30B}) is already contained in $\mathcal{L}_2$ in flat spacetime, up to a total derivative, so that, in this framework, the GSU2P does not contain terms built exclusively with three derivatives that are linearly independent of $\mathcal{L}_2$. The conclusion is, nonetheless, completely different in curved spacetime.

\subsection{Covariantization}

The minimal covariantization scheme applied to the suitable combination $\partial_\mu (2 \tilde{J}^\mu_1 + \tilde{J}^\mu_2)$ of terms in Eq. (\ref{td30}) leads to
 \begin{eqnarray}
\nabla_\mu (2 \tilde{J}^\mu_1 + \tilde{J}^\mu_2) &=& \nonumber \\
&& (... \in \mathcal{L}_2) \nonumber \\
&& +  \frac{3}{8} (- 2 \mathcal{L}_5^1 + \mathcal{L}_5^3) \nonumber \\
&& + \frac{1}{2} A^{\nu a} R^\sigma_{\ \ \nu \rho \mu} A_\sigma^b A_{\alpha \beta}^c \epsilon^{\mu \rho \alpha \beta} \epsilon_{abc} \,. \nonumber \\
&&
\end{eqnarray}
The Lagrangian in Eq. (\ref{L30B}) is, therefore, not redundant against $\mathcal{L}_2$ in curved spacetime.  As a remainder of this fact, we will dismiss $- 2 \mathcal{L}_5^1 + \mathcal{L}_5^3$ in favour of $\frac{1}{2} A^{\nu a} R^\sigma_{\ \ \nu \rho \mu} A_\sigma^b A_{\alpha \beta}^c \epsilon^{\mu \rho \alpha \beta} \epsilon_{abc}$.  We conclude then that the reconstructed GSU2P exhibits the following Lagrangian built from just three derivatives:
\begin{empheq}[box=\widefcolorbox]{align}
\tilde{\mathcal{L}}_{5,0} = A^{\nu a} R^\sigma_{\ \ \nu \rho \mu} A_\sigma^b \tilde{A}^{\mu \rho c} \epsilon_{abc} \,. \label{L50}
\end{empheq}

\subsection{The decoupling limit}

Since the Lagrangian given in the previous expression vanishes in the decoupling limit $A_\mu^a \rightarrow \nabla_\mu \pi^a$, because of the antisymmetry of $\tilde{A}_{\mu \nu}^a$, it is free of the Ostrogradski instability.

\section{Comparison with the ``old'' GSU2P} \label{compold}

The old GSU2P, formulated in Ref. \cite{Allys:2016kbq}, is described by the following Lagrangian:
\begin{equation}
\mathcal{L}^{\rm old} = \mathcal{L}_2^{\rm old} + \sum_{i = 1}^3 \alpha_i \mathcal{L}_4^{i, \rm{old}} + \sum_{i = i}^4 \beta_i \mathcal{L}_{\rm Curv}^{i, \rm{old}} \,,
\end{equation}
where the $\alpha_i$ and $\beta_i$ are dimensionful arbitrary constants, $\mathcal{L}_2^{\rm old} \equiv \mathcal{L}_2^{\rm old} (A_{\mu \nu}^a, A_\mu^a)$ is an arbitrary function of $A_{\mu \nu}^a$ and $A_\mu^a$, and
\begin{eqnarray}
\mathcal{L}_4^{1, \rm{old}} &=& (A_b \cdot A^b) [S^{\mu a}_\mu S^\nu_{\nu a} - S^{\mu a}_\nu S^\nu_{\mu a} - R (A^a \cdot A_a)] \nonumber \\
&& + 2 (A_a \cdot A_b) [S^{\mu a}_\mu S^{\nu b}_\nu - S^{\mu a}_\nu S^{\nu b}_\mu - R (A^a \cdot A^b)] \,, \nonumber \\
\mathcal{L}_4^{2, \rm{old}} &=& (A_a \cdot A_b) [S^{\mu a}_\mu S^{\nu b}_\nu - S^{\mu a}_\nu S^{\nu b}_\mu - R (A^a \cdot A^b)]  \nonumber \\
&& + A_\mu^a A_\nu^b [S^{\mu \alpha}_a S^\nu_{\alpha b} - S^{\mu \alpha}_b S^\nu_{\alpha a} \nonumber \\
&& + 2 A^{\mu \alpha}_a S^\nu_{\alpha b} - 2 A^{\mu \alpha}_b S^\nu_{\alpha a} + 2 A_{\rho a} A_{\sigma b} R^{\mu \nu \rho \sigma}] \,, \nonumber \\
\mathcal{L}_4^{3, \rm{old}} &=& A^\mu_a \tilde{A}_{\mu \sigma}^b S^{\sigma \nu a} A_{\nu b} \,, \nonumber \\
\mathcal{L}_{\rm Curv}^{1, \rm{old}} &=& G_{\mu \nu} A^{\mu a} A^\nu_a \,, \nonumber \\
\mathcal{L}_{\rm Curv}^{2, \rm{old}} &=& L_{\mu \nu \rho \sigma} A^{\mu \nu a} A^{\rho \sigma}_a \,, \nonumber \\
\mathcal{L}_{\rm Curv}^{3, \rm{old}} &=& L_{\mu \nu \rho \sigma} A^{\mu \nu a} A^{\rho b} A^{\sigma c} \epsilon_{abc} \,, \nonumber \\
\mathcal{L}_{\rm Curv}^{4, \rm{old}} &=& L_{\mu \nu \rho \sigma} A^{\mu a} A^{\nu b} A^\rho_a A^\sigma_b \,, \label{Lold}
\end{eqnarray}
where $L_{\mu \nu \rho \sigma} \equiv \frac{1}{2} \epsilon_{\mu \nu \alpha \beta} \epsilon_{\rho \sigma \gamma \delta} R^{\alpha \beta \gamma \delta}$ is the double dual of the Riemann tensor.  This old theory was built following the same steps that we followed here except for three aspects:
\begin{enumerate}
\item All the Lagrangian building blocks were constructed employing the full $\partial_\mu A_\nu^a$ instead of splitting it into its symmetric $S_{\mu \nu}^a$ and antisymmetric $A_{\mu \nu}^a$ parts.  This, of course, produced a lot more blocks (and a lot more work) than needed, many linear combinations of them already included in $\mathcal{L}_2$.
\item Only the primary constraint-enforcing relation was considered.  As was shown in Refs. \cite{ErrastiDiez:2019ttn,ErrastiDiez:2019trb}, this is not enough to remove the Ostrogradski ghost.
\item Many terms were dismissed by employing total derivatives already at the flat spacetime level, which led to a loss of several terms that exists only in curved spacetime, including the beyond SU(2) Proca ones.  Moreover, most of the total derivatives employed do not satisfy the secondary Hessian constraint.
\end{enumerate}
The application of this theory to inflation and dark energy was investigated in Refs. \cite{Rodriguez:2017ckc,Rodriguez:2017wkg}, and the stability analysis of the same was performed in Ref. \cite{Gomez:2019tbj},  so we wonder how the results of these works could change in the light of the new theory presented in this paper.

As can be seen, our $\hat{\mathcal{L}}_4^{1,h}$ in Eq. (\ref{L41h}) is identical to $\mathcal{L}_4^{1, \rm{old}}$, this being one of the reasons of the second change in basis elements in Section \ref{chba}.  Examined from the viewpoint of the reconstructed GSU2P [see Eqs. (\ref{L42})-(\ref{L42LPa})], $\mathcal{L}_4^{1, \rm{old}}$ can also be written as
\begin{equation}
\mathcal{L}_4^{1, \rm{old}} = \mathcal{L}_{4,2}^1 - \mathcal{L}_{4,2}^4 \,.
\end{equation}
Thus, we conclude that $\mathcal{L}_4^{1, \rm{old}}$ is free of the Ostrogradski ghost (at least in flat spacetime).  

Now, $\mathcal{L}_4^{2, \rm{old}}$ was shown in Ref. \cite{ErrastiDiez:2019trb} not to satisfy the secondary Hessian constraint and, so, as an example of the ghost instabilities that plagued the old GSU2P.  Nevertheless, a bit of algebra shows us that 
\begin{eqnarray}
\mathcal{L}_4^{2, \rm{old}} - 2 \partial_\mu \tilde{J}^\mu_2 &=& (... \in \mathcal{L}_2) \nonumber \\
&& + \frac{1}{12} \mathcal{L}_{4,2}^1 - \frac{1}{3} \mathcal{L}_{4,2}^2 \nonumber \\
&& - \frac{5}{9} \partial_\mu (\tilde{J}^\mu_3 + 3 \tilde{J}^\mu_2) + \frac{1}{9} \partial_\mu (2 \tilde{J}^\mu_3 - 3 \tilde{J}^\mu_1) \,, \nonumber \\
&&
\end{eqnarray}
at the flat space-time level, where the quantities in this expression, except for $\mathcal{L}_4^{2, \rm{old}}$, are those of Section \ref{42}.  Thus, although neither $\mathcal{L}_4^{2, \rm{old}}$ is healthy, nor $\partial_\mu \tilde{J}^\mu_2$ is, the combination $\mathcal{L}_4^{2, \rm{old}} - 2 \partial_\mu \tilde{J}^\mu_2$ satisfies the secondary constraint-enforcing relation, and, therefore, all the physics extracted from the unhealthy curved space-time version of $\mathcal{L}_4^{2, \rm{old}}$, for instance, in Ref. \cite{Gomez:2019tbj}, is equivalent to that extracted from the healthy $\mathcal{L}_4^{2, \rm{old}} - 2 \nabla_\mu \tilde{J}^\mu_2$.

Something similar occurs for $\mathcal{L}_4^{3, \rm{old}}$:
\begin{eqnarray}
\mathcal{L}_4^{3, \rm{old}} + 2 \partial_\mu \tilde{J}^\mu_4 &=& (... \in \mathcal{L}_2) \nonumber \\
&& + \frac{1}{2} (\tilde{\mathcal{L}}_{4,2}^1 + 4 \partial_\mu \tilde{J}^\mu_6) \,,
\end{eqnarray}
at the flat space-time level, so although neither $\mathcal{L}_4^{3, \rm{old}}$ nor $\partial_\mu \tilde{J}^\mu_4$ are healthy, the combination $\mathcal{L}_4^{3, \rm{old}} + 2 \partial_\mu \tilde{J}^\mu_4$ is, and, therefore, all the physics extracted from the unhealthy curved space-time version of $\mathcal{L}_4^{3, \rm{old}}$ is equivalent to that extracted from the healthy $\mathcal{L}_4^{3, \rm{old}} + 2 \nabla_\mu \tilde{J}^\mu_4$.

Now, as can be seen in Eq. (\ref{L42LPb}), there exist only two parity-violating terms in flat spacetime in the reconstructed GSU2P.  Then, why is it that in the old GSU2P there exists only one?  Leaving aside the fact that $\tilde{\mathcal{L}}_{4,2}^1$ might be unhealthy in its decoupling limit, the reason lies in a small mistake in the conditions of Eq. (37) in Ref. \cite{Allys:2016kbq} to make the primary constraint-enforcing relation vanish that prevented the authors of that work from finding a second parity-violating Lagrangian piece.

Finally, among the $\mathcal{L}_{\rm Curv}^{\rm old}$ of Eq. (\ref{Lold}), the only one that appears in the reconstructed theory is $\mathcal{L}_{\rm Curv}^{1, {\rm old}}$, which is exactly the same as our $\mathcal{L}_{4,0}$ of Eq. (\ref{L20}).  The other $\mathcal{L}_{\rm Curv}^{\rm old}$ were just postulated, as they are obviously healthy because of the divergenceless nature of $L_{\mu \nu \rho \sigma}$.  We could have postulated them as well in the reconstructed GSU2P, but we would rather not do it.  This is because we expect them to naturally appear in the theory when more than six space-time indices are considered in the Lagrangian building blocks without contractions.

\section{Comparison with the generalized Proca theory} \label{compgpt}

Finding the beyond GSU2P in Section \ref{42} required determining the values of the constants $\tilde{a}$ and $\tilde{b}$ in Eq. (\ref{L42WDLC}).  We could have followed the standard procedure of finding out the kinetic matrix of its decoupling limit and making it degenerate \cite{Langlois:2015cwa,Ganz:2020skf}.  However, we followed an alternative route based on the fact that the GSU2P stripped of the internal group indices must be contained in the generalized Proca theory.  Indeed, the other reason why we performed the second change in basis elements in Section \ref{chba} is that $\hat{\mathcal{L}}_4^{1,h}$ stripped of the internal group indices is nothing else than $\mathcal{L}_4$ of the generalized Proca theory (see Ref. \cite{Jimenez:2016isa}):
\begin{equation}
\mathcal{L}_4 = G_4(X) R + \frac{G_{4,X}(X)}{4} (S^\mu_\mu S^\nu_\nu - S^\mu_\nu S^\nu_\mu) \,,
\end{equation}
for $G_4(X) = - 3X^2$.  Then, what about the other Lagrangian pieces that make $\mathcal{L}_{4,0}$ and $\mathcal{L}_{4,2}$?  First of all, $\mathcal{L}_{4,0}$ stripped of the internal group indices is just $\mathcal{L}_4$, up to a total derivative, with $G_4(X) = X$.  In contrast, $\mathcal{L}_{4,2}^2$ 
and $\tilde{\mathcal{L}}_{4,2}^1$ 
reduce to zero when stripped of the internal group indices.  Regarding $\mathcal{L}_{4,2}^5$ and $\mathcal{L}_{4,2}^6$ without internal group indices, they are just healthy extensions of $G_{\mu \nu} A^\mu A^\nu$ that were not recognized in Ref. \cite{Allys:2016kbq}.  Finally, $\tilde{\mathcal{L}}_{4,2}^2$, $\tilde{\mathcal{L}}_{4,2}^3$, and $\tilde{\mathcal{L}}_{4,2}^4$, stripped of their internal group indices, reduce, up to total derivatives, to $A_\beta \tilde{A}^{\beta \alpha} S_{\alpha \rho} A^\rho$, which was shown in Refs. \cite{Rodriguez:2017ckc,Kimura:2016rzw} to be part of $\mathcal{L}_2$ up to a total derivative.  To end up, the only parity-violating terms in the generalized Proca theory belong to $\mathcal{L}_2$ \cite{Allys:2016jaq}, so $\tilde{\mathcal{L}}_{5,0}$ stripped of its internal group indices should be either zero, a total derivative, or contained in $\mathcal{L}_2$;  in fact, observing Eq. (\ref{L50}), the first alternative is the correct one.

\section{Conclusions} \label{conc}

GSU2P and beyond GSU2P are described by the Lagrangians in Eq. (\ref{L20}), Eqs. (\ref{L42})-(\ref{L42LPb}), and Eq. (\ref{L50}).  The theory has been written so as to make it explicit which Lagrangian pieces exist only in curved spacetime and which ones exist even in flat spacetime;  indeed, from the twelve Lagrangian pieces that compose the theory, only four, $\mathcal{L}_{4,2}^1, \mathcal{L}_{4,2}^2, 
\tilde{\mathcal{L}}_{4,2}^1$, 
and $\tilde{\mathcal{L}}_{4,2}^2$, survive in flat spacetime.  The nature of some of the Lagrangian pieces is purely non-Abelian -- i.e., they vanish when stripped of their internal group indices --  specifically, $\mathcal{L}_{4,2}^2, 
\tilde{\mathcal{L}}_{4,2}^1$, 
and $\tilde{\mathcal{L}}_{5,0}$ belong to this subset.  It is worthwhile mentioning that $\tilde{\mathcal{L}}_{4,2}^2$ is the parity-violating version of $\mathcal{L}_{4,2}^2$ as can be easily observed.  On the other hand, the theory is diffeomorphism invariant, so that the energy and momentum are locally conserved \cite{Misner:1974qy}.

Much remains to be done in the exploration of this theory as a candidate of an effective theory for the gravitational interaction.  First of all, it is not clear yet whether the decoupling limits of the beyond GSU2P terms  
as well as that of $\tilde{\mathcal{L}}_{4,2}^1$ 
are actually healthy\footnote{It is unlikely that the decoupling limits of the beyond GSU2P terms are unhealthy:  there is actually no reason to believe that healthy beyond extensions do exist for the Horndeski theory and the generalized Proca theory but do not for the GSU2P.  In contrast, there is no clue regarding the healthiness of $\tilde{\mathcal{L}}_{4,2}^1$, this term being of a purely non-Abelian nature.}.
Other self-consistency issues must be addressed, such as the possible existence of ghosts (other than the Ostrogradski one) and Laplacian instabilities, as a follow-up of the work in Ref. \cite{Gomez:2019tbj}, the generalization of the constraint algebra to curved spacetime \cite{ErrastiDiez:2020dux,Heidari:2020cil}, the analysis of the causal structure \cite{Hawking:1973uf}, and the calculation of the cutoff scale of the theory and its comparison with the GW170817 event frequency \cite{deRham:2018red} (to see whether the bound on the gravitational waves speed applies to GSU2P \footnote{How the gravitational wave speed bound affects the generalized Proca theory was investigated in Ref. \cite{Baker:2017hug}.}).  We might as well construct an extended version of this theory, considering all the possibilities to degenerate the kinetic matrix in curved spacetime, as was done for the generalized Proca theory in Refs. \cite{Kimura:2016rzw,deRham:2020yet}.  The theory must, of course, be put under test against observations;  in this regard, determining whether there exists a screening mechanism at Solar System scales, as was studied in Ref. \cite{DeFelice:2016cri} for the generalized Proca theory, is a crucial aspect.  Of course, the cosmological and astrophysical implications must be properly studied both at the background (see, for instance, Ref. \cite{Rodriguez:2017wkg}) and at the perturbative level (see, for instance, Refs. \cite{Dimopoulos:2008yv,Gomez:2013xza,Almeida:2014ava}).  We finish this paper by reminding the readers and ourselves of one important message given to us by Misner, Thorne, and Wheeler in their marvelous treatise on gravitation \cite{Misner:1974qy}:  {\it ``To be complete, a theory of gravity must be capable of analyzing from `first principles' the outcome of every experiment of interest.  It must therefore mesh with and incorporate a consistent set of laws for electromagnetism, quantum mechanics, and all other physics.''}  There is a long road in this direction ahead of us that we hope we will travel.

\section*{Acknowledgments} 

We appreciate all the discussions and exchange of ideas we had with Juan Camilo Garnica Aguirre and Carlos Mauricio Nieto Guerrero 
which helped a lot in the development of this work.  The work presented here was supported by the following grants: Colciencias-Deutscher Akademischer Austauschdienst Grant No. 110278258747 RC-774-2017, Vicerrector\'{\i}a de Ciencia, Tecnolog\'{\i}a, e Innovaci\'on - Universidad Antonio Nari\~no Grant No. 2019248, and Direcci\'on de Investigaci\'on y Extensi\'on de la Facultad de Ciencias - Universidad Industrial de Santander Grant No. 2460.  A.G.C. was supported by Beca de Inicio Postdoctoral REXE RA No. 315-3269-2020 Universidad de Valpara\'{\i}so. L.G.G. was supported by the postdoctoral scholarship No.  2020000102 Vicerrector\'{\i}a de Investigaci\'on y Extensi\'on - Universidad Industrial de Santander. Some calculations were cross-checked with the {\it Mathematica} package xAct (see www.xact.es).

\bibliography{Bibli.bib} 

\end{document}